\documentclass[aps,prl,twocolumn,superscriptaddress,floatfix,nofootinbib]{revtex4-2}
\usepackage[T1]{fontenc}
\usepackage[utf8]{inputenc}
\usepackage{amsmath,amssymb,bm,mathtools}
\usepackage{graphicx,booktabs}
\usepackage{xcolor}
\usepackage[colorlinks=true,linkcolor=blue!50!black,citecolor=blue!50!black,urlcolor=blue!50!black]{hyperref}
\graphicspath{{figures/}}
\newcommand{\bk}{\bm{k}}
\newcommand{\avg}[1]{\langle #1\rangle_{\bk}}
\newcommand{\Nf}{N_f}
\newcommand{\dd}{\mathrm d}
\newcommand{\Impart}{\operatorname{Im}}
\newcommand{\Repart}{\operatorname{Re}}
\newcommand{\hf}{\hat f}
\newcommand{\Gloc}{G^{\rm loc}}
\newcommand{\Sig}{\Sigma_f}

\begin{document}

\title{Quantum-geometry-tuned Mott transition and emergent heavy fermions in a topological flat band}

\author{Ya-Hui Zhang}
\affiliation{William H. Miller III Department of Physics and Astronomy, Johns Hopkins University, Baltimore, Maryland 21218, USA}

\date{\today}

\begin{abstract}
A flat band with a nonzero Chern number has no localized Wannier orbitals, which raises the question of whether Mott physics is still possible. We place a Hubbard interaction $U$ on coherent orbitals, which are localized but overcomplete, and develop a single-site dynamical mean-field theory (DMFT) for this model. Within the flat band alone, DMFT reproduces the experimentally observed cascade and inverse-compressibility sawtooth without invoking Dirac revival or remote-band electrons. At nonzero integer filling we find an insulator-to-metal transition tuned by $s^2$, the fraction of the Brillouin zone that carries the Berry curvature. For $\Nf=8$ flavors the Mott transition at $\nu=-2$ occurs at $s^2_c\approx0.025$, where the top of the lower Hubbard band reaches the Fermi level and the local moments are Kondo screened, forming a heavy Fermi liquid (HFL). The HFL extends to the hole-doped side, $\nu=-2-x$, but above the small Kondo temperature $T_K$ the state is an incoherent bad metal without a Fermi surface. At low temperature, an anti-Hund's coupling $J_A\sim0.1U$ replaces the HFL with a pseudogap metal that has a small Fermi surface and is dominated by spin-polaron excitations at small $x$. We propose this pseudogap metal as the parent state of superconductivity in TBG. The framework may also be applied to other topological bands, such as those of twisted WSe$_2$.
\end{abstract}

\maketitle

\emph{Introduction.}---Mott physics is usually formulated as a Hubbard model on localized Wannier orbitals, which is impossible for a topological band~\cite{Brouder2007}. Yet magic-angle twisted bilayer graphene (MATBG)~\cite{Bistritzer2011}, whose flat bands are topological~\cite{Po2018,Song2019,Tarnopolsky2019}, shows the phenomenology of Mott physics: correlated insulators and superconductivity~\cite{Cao2018a,Cao2018b,Yankowitz2019,Lu2019,Cao2021}, local moments with a large entropy~\cite{Saito2021,Rozen2021}, a cascade of transitions at every integer~\cite{Zondiner2020,Wong2020} and a pseudogap in tunneling~\cite{Oh2021,Park2026,Kim2026}. In the chiral limit its bands are ideal Chern bands close to a lowest Landau level~\cite{Tarnopolsky2019,Wang2021,Popov2021,Ledwith2022}, the kind of band that hosts fractional Chern insulators~\cite{Tang2011,Sun2011,Neupert2011,Sheng2011,Regnault2011,Parameswaran2013,Cai2023,Zeng2023,Park2023,Xu2023,Lu2024}, and Hartree--Fock in this Chern basis accounts for the symmetry-broken insulators at integer fillings~\cite{Kang2019,Bultinck2020,Lian2021,Kwan2021,Wagner2022,Hong2022,Wang2023,Repellin2020}.

In MATBG the Wannier obstruction is fragile~\cite{Po2018,Po2019,Ahn2019,Song2019} and can be removed by adding remote bands, which gives the topological heavy-fermion model~\cite{Song2022,Shi2022,Calugaru2023}: localized $f$ orbitals hybridize with itinerant $c$ electrons of the remote bands. This model has motivated Kondo-lattice~\cite{Chou2023,Hu2023,Hu2023b,Zhou2024,Lau2023} and DMFT~\cite{Datta2023,Rai2024,Youn2024,Calugaru2025} studies. Once the model is projected onto the flat bands, the itinerant electrons are gone, and it is not obvious whether Mott and heavy-fermion physics survive. Recent works discuss Mott physics in TBG, including a Mott semimetal at charge neutrality~\cite{Ledwith2025,LedwithTrion2025,Shan2026,ZhaoMixed2025,ZhaoAncilla2025,ZhouWang2026} and results equivalent to the Hubbard-I~\cite{Ledwith2025,ZhaoMixed2025,ZhaoAncilla2025,HuSong2025,Calugaru2025} and Hubbard-III~\cite{Vituri2026,Wei2026,Hu2026,Nosov2026} approximations. Most of these descriptions focus on the single-electron self-energy and do not fully include the localized spin moments. They are valid if the Hubbard bands decouple from the local moments or couple to them only weakly. A recent work by the author~\cite{Zhang2026} argues that the charge carriers couple to the local moments through a Kondo exchange $J_K\sim U$, so that the Hubbard-I and Hubbard-III approximations may not be valid. The question then arises: what framework can capture both the nontrivial momentum-selective Hubbard bands and the local moments when the two are strongly coupled?

We address this question by considering a Hubbard model on coherent orbitals, obtained by projecting a localized orbital onto the flat band, similar to the coherent states of the lowest Landau level~\cite{Perelomov1971,Bacry1975} (see also~\cite{Okuma2025,Okuma2026}). They are localized, but overcomplete and non-orthogonal, with an overlap set by $s^2$, the fraction of the Brillouin zone that carries the Berry curvature. We develop a dynamical mean-field theory (DMFT)~\cite{Georges1996} built on them with a bath that comes entirely from this overlap and is of order $s^2$. Our DMFT calculations reproduce the cascade in spectroscopy~\cite{Wong2020,Choi2021,Xiao2025} and the sawtooth in the inverse compressibility~\cite{Tomarken2019,Zondiner2020,Rozen2021,Pierce2021,Yu2022,HuQ2025,ZhangAndrei2026} observed in TBG, including inverse-compressibility maxima shifted away from the integer fillings~\cite{HuQ2025}. We therefore suggest that these phenomena may not require itinerant electrons from remote bands or Dirac revival~\cite{Zondiner2020}. Because it keeps only the flat band, our model is numerically cheaper than previous DMFT studies that include the remote bands~\cite{Datta2023,Rai2024,Youn2024,Calugaru2025}, and its essential physics is easier to extract.

\begin{figure}[t]
 \includegraphics[height=0.38\columnwidth]{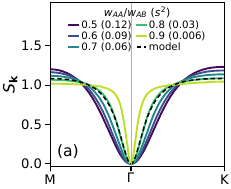}\hfill
 \includegraphics[height=0.38\columnwidth]{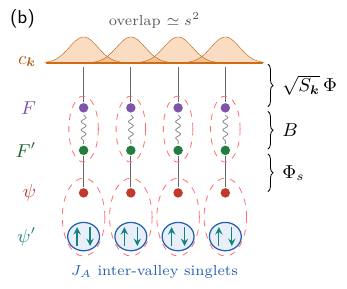}\\[2pt]
 \includegraphics[width=\columnwidth]{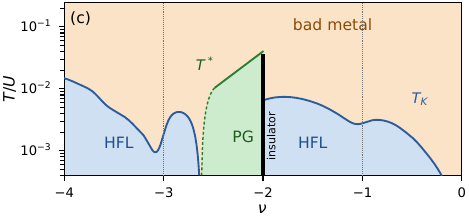}
 \caption{(a) Weight $S_{\bk}$ of the coherent orbital of MATBG for several $w_{AA}/w_{AB}$, with $s^2$ in parentheses. The dashed line is the Lorentzian texture at $s^2=0.03$. (b) The ancilla theory of Eqs.~\eqref{eq:chain} and \eqref{eq:pgchain}. The band electron $c_{\bk}$ (lobes, the coherent orbitals) couples to the orthogonal fermion $F$, $F$ couples to the spinon $F'$ that carries the moment~\cite{ZhangSachdev2020,Zhang2026}, and $F'$ couples to the spin polaron $\psi$. The inter-valley singlets from $J_A$ are denoted $\psi'$. (c) Phase diagram without flavor order for $\Nf=8$, $s^2=0.05$ and $J_A^d=0.1U$. On cooling, the bad metal becomes a heavy Fermi liquid (HFL) below the Kondo temperature $T_K$ or, at $\nu=-2-x$, the pseudogap metal (PG) below $T^*$. The solid green line is $T^*$ from DMFT (Appendix~\ref{app:jad}), and its dashed continuation is schematic. $T_K$ is from DMFT without $J_A$. The thick line marks the insulator at $\nu=-2$.}
 \label{fig:intro}
\end{figure}

A perturbative expansion in $s^2$ recovers the Hubbard-I and Hubbard-III approximations and leads to an intuitive emergent heavy-fermion picture: the band electron hybridizes with an orthogonal fermion $F$, which in turn couples to the local moment through a Kondo exchange $J_K\sim U$. Self-consistent DMFT departs qualitatively from the Hubbard-I and Hubbard-III approximations for $s^2\gtrsim0.02$ and finds a Mott transition at non-zero integer filling, beyond which the local moments are Kondo screened. For $s^2=0.05$, a realistic value for TBG [Fig.~\ref{fig:intro}(a)], the state below the Kondo temperature $T_K\lesssim0.01U$ is a heavy Fermi liquid (HFL) over the whole range of doping. Above $T_K$ it is a bad metal without a Fermi surface, and its self-energy is fit by the ancilla theory with $F$ and the spinon $F'$ [Fig.~\ref{fig:intro}(b)]. At $\nu=-2-x$, a small anti-Hund's coupling $J_A\sim0.1U$ replaces the HFL with a pseudogap metal, which becomes a second Fermi liquid (sFL)~\cite{ZhangSachdev2020,ZhaoZhang2025,OhYangZhang2025,ZhaoNickelate2026} with a small Fermi surface below a low coherence temperature $T_{\rm coh}\sim U/800$. To describe this sFL phase analytically, we need to introduce another pair of ancilla fermions $\psi$ and $\psi'$. Our theory offers a description of the normal states of TBG [Fig.~\ref{fig:intro}(c)] and establishes the possibility of a Mott transition tuned by quantum geometry instead of kinetic energy.

\emph{Model and DMFT.}---We consider $\Nf$ flavors, each with an exactly flat band of Chern number $C=\pm 1$. A localized orbital on each site $\bm R_i$, with the form factor $\mu_f(\bk)$ on the Bloch states and the weight $w_f=|\mu_f|^2$, whose Brillouin-zone average is $W_f$, defines the coherent orbital
\begin{equation}
 \hf_{i}=\frac1{\sqrt{N_M}}\sum_{\bk}e^{i\bk\cdot\bm R_i}\,\frac{\mu_f(\bk)}{\sqrt{W_f}}\,c_{\bk},\qquad
 S_{\bk}=\frac{w_f(\bk)}{W_f},
 \label{eq:orb}
\end{equation}
which is normalized, $\avg{S_{\bk}}=1$, but overlaps with its neighbors: $\{\hf_i,\hf^\dagger_j\}=S_{ij}=N_M^{-1}\sum_{\bk}e^{i\bk\cdot(\bm R_i-\bm R_j)}S_{\bk}$, with $S_{ii}=1$. In a Chern band $\mu_f$ vanishes at a point $\bk_0=0$ around which its phase winds~\cite{Perelomov1971,Bacry1975}. Near it $w_f(\bk)\simeq k^2/(k^2+k_*^2)$, and $s^2=\pi k_*^2/\Omega_{\rm BZ}$ is the fraction of the zone in which $w_f<1/2$. The total overlap of one orbital with all others is $\eta_S^2=\sum_{j\neq i}|S_{ij}|^2=\avg{S_{\bk}^2}-1\simeq s^2$. A lowest Landau level with spatially varying magnetic field realizes this band, with $s^2\simeq\bar B/B_0$ for $B_0\gg\bar B$, the ratio of the mean field to that on a site (Appendix~\ref{app:coherent}). In MATBG, $\hf$ can be the $f$ orbital of the topological heavy-fermion model~\cite{Song2022} projected onto the flat bands, or be built directly from the continuum flat bands. Then $s^2$ falls with the ratio $w_{AA}/w_{AB}$ of the interlayer tunnelings, from $s^2=0.09$  to $s^2=0.03$ when $w_{AA}/w_{AB}$ changes from $0.6$ to $0.8$ [Fig.~\ref{fig:intro}(a)]. We put only a Hubbard interaction $U$ on these orbitals,
\begin{equation}
 H=-\mu\sum_{\bk}c^\dagger_{\bk}c_{\bk}
 +\frac U2\sum_i\Big(\hat n_{fi}-\frac{\Nf}2\Big)^2,
 \label{eq:ham}
\end{equation}
with $\hat n_{fi}=\sum_\alpha\hf^\dagger_{i\alpha}\hf_{i\alpha}$. Here $n_c$ is the band filling per site, and $\nu=n_c-\Nf/2$. DMFT takes the self-energy to be local in the $\hat f$ orbitals: $\Sig(z)\sum_i\hf^\dagger_i\hf_i=\Sig(z)\sum_{\bk}S_{\bk}c^\dagger_{\bk}c_{\bk}$. With $\zeta=z+\mu$, the band electron and the local orbital then propagate as
\begin{equation}
 G_c(\bk,z)=\frac1{\zeta-S_{\bk}\Sig(z)},\qquad
 \Gloc_f(z)=\avg{S_{\bk}G_c(\bk,z)} .
 \label{eq:dmft}
\end{equation}
The impurity site of DMFT is one $\hat f$ orbital with the same Hubbard interaction and an additional bath $\Delta_f(z)$. Requiring its Green function $G_f^{\rm imp}=(\zeta-\Delta_f-\Sig)^{-1}$ to equal $\Gloc_f$ gives the self-consistency $\Delta_f=\zeta-\Sig-1/\Gloc_f$. The bath vanishes for orthonormal orbitals, $S_{\bk}\equiv1$: it comes only from the mixed states, $0<S_{\bk}<1$. We solve the impurity site by hybridization-expansion continuous-time quantum Monte Carlo (CT-HYB) in its segment form~\cite{Werner2006,Gull2011,Kavokine2025,Parcollet2015} for $U/3200\le T\le U$. All results are for $\Nf=8$ and $S_{\bk}=W_f^{-1}k^2/(k^2+k_*^2)$, regularized on the triangular lattice as in Eq.~\eqref{eq:wlat}. 

\begin{figure}[t]
 \includegraphics[width=\columnwidth]{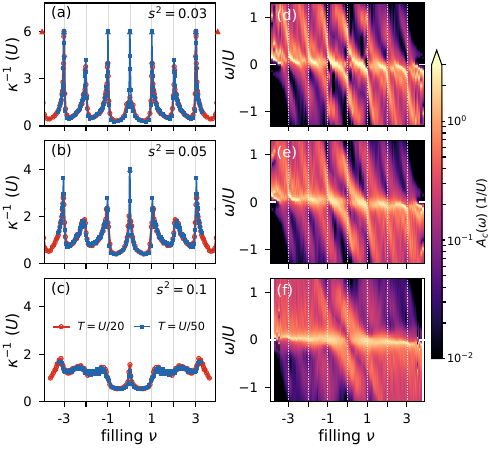}
 \caption{Cascade within the flat band ($\Nf=8$) for $s^2=0.03$, $0.05$ and $0.10$. (a--c) Inverse compressibility $\kappa^{-1}=\dd\mu/\dd n_c$ at $T=U/20$ (red) and $U/50$ (blue).  (d--f) Local spectral function $A_c(\omega)$ per flavor versus filling at $T=U/20$  on a logarithmic color scale. $\nu>0$ follows from particle-hole symmetry. }
 \label{fig:cascade}
\end{figure}

\emph{Expansion in $s^2$ and emergent heavy fermions.}---At $s^2=0$ every site is an isolated SU($\Nf$) atom. With $\mu$ in one of its charge gaps it has an integer charge $n_f$. With $q=n_f/\Nf$ the atomic self-energy has a single pole,
\begin{equation}
 \Sigma_{\rm at}(z)=\Sigma_\infty+\frac{\Phi^2}{z-\epsilon_F},\qquad \Phi^2=U^2q(1-q).
 \label{eq:hub1}
\end{equation}
The above self-energy can be generated by a mean-field theory in the ancilla framework~\cite{ZhangSachdev2020,ZhaoMixed2025,ZhaoAncilla2025,Zhang2026}, where the physical electron $c_{\bk}$ couples to an ancilla fermion $F$ at the level $\epsilon_F$ with the amplitude $\sqrt{S_{\bk}}\,\Phi$. To linear order in $s^2$ we simply use $\Sigma_{\rm at}(z)$ to generate the bath $\Delta_f(z)$, whose spectral density is $\rho_\Delta(\omega)=s^2|\omega+\mu|$. From this bath we obtain a correction to $\Sig$ through the self-energy $\Sigma_F$ of $F$,
\begin{equation}
 \Sig(z)=\Sigma_\infty+\frac{\Phi^2}{z-\epsilon_F-\Sigma_F(z)},
 \label{eq:dysonF}
\end{equation}
which at this order is
\begin{equation}
 \begin{gathered}
 \Sigma_F(z)=(\Nf+1)\Delta_f(z),\\
 \Gamma_F(\omega)\equiv-\Impart\Sigma_F(\omega+i0^+)=(\Nf+1)\,\pi s^2|\omega+\mu|,
 \end{gathered}
 \label{eq:hub3}
\end{equation}
where the static part of $\Delta_f$ is absorbed in the chemical potential, $\Gamma_F$ is the scattering rate of $F$, and the result matches Hubbard's third approximation~\cite{Hubbard1964,Wei2026}. Away from a non-zero integer filling, on its hole side, the Fermi level lies in the band of $F$, and the scattering rate $\Gamma_F(0)=(\Nf+1)\pi s^2|\mu|$ is large, since $|\mu|\simeq|\nu|U$. This scattering rate can be interpreted as the Born approximation of a Kondo exchange between $F$ and the SU($\Nf$) moment $S^{\alpha\beta}$~\cite{Zhang2026},
\begin{equation}
 H_K=\frac{J_K}{N_M}\sum_{\bk\bk'}\phi_{\bk}\phi_{\bk'}F^\dagger_{\bk\alpha}S^{\alpha\beta}F_{\bk'\beta},\qquad J_K=\frac{U\Phi^2}{E_+E_-}\sim U,
 \label{eq:jk}
\end{equation}
with $E_\pm$ the energies of adding and removing one electron and $\phi_{\bk}\propto1-S_{\bk}$ a form factor (Appendix~\ref{app:kondo}). The effective dimensionless coupling is $g=\Nf J_K\rho_F(0)=\Nf U\rho_\Delta(0)/(E_+E_-)$, with $\rho_F=\rho_\Delta/\Phi^2$ the density of states of $F$, of order $s^2$. The $O(s^2)$ expansion above is therefore the leading order in $g$, but $g$ grows under renormalization, $dg/d\ln D=-g^2$, and reaches strong coupling at the Kondo temperature $T_K\simeq De^{-1/g}$, where the cutoff $D\approx0.08U$ is the width of the band of $F$ (Appendix~\ref{app:tk}). Perturbation theory breaks down when $T$ reaches $T_K$, and there we turn to the full DMFT calculation.

\begin{figure}[t]
 \includegraphics[width=\columnwidth]{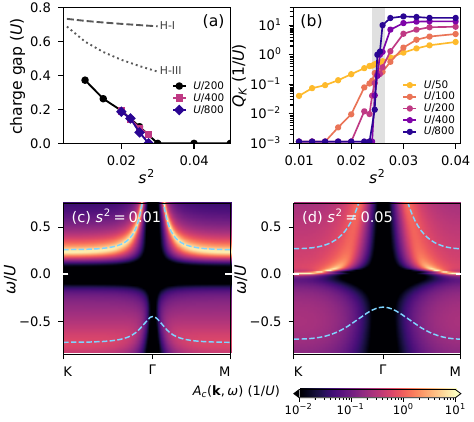}
 \caption{Mott transition at $\nu=-2$ ($\Nf=8$). (a) Charge gap, the width of the plateau of $n_c(\mu)$, versus $s^2$. The dashed and dotted lines are Hubbard-I and Hubbard-III. (b) Weight at the Fermi level at the zone corner, $Q_K=-\beta G_c(K,\beta/2)/\pi$, at the center of the plateau, without continuation. Values below $10^{-3}$ are drawn at the bottom of the axis, and the shaded band marks the transition. (c,d) $A_c(\bk,\omega)$ at $T=U/400$ for $s^2=0.01$ and $0.05$, the latter below $T_K$. The dashed lines are the Hubbard-I bands of Eq.~\eqref{eq:hub1}.}
 \label{fig:mott}
\end{figure}

\emph{Cascade.}---We now turn to DMFT results. First we sweep over $\nu$ and show the inverse compressibility and the local spectral function $A_c(\omega)$ in Fig.~\ref{fig:cascade}. At $s^2=0.03$ the inverse compressibility peaks at every integer and then decays on its hole side roughly as $1/\sqrt{\delta\nu}$ in the distance $\delta\nu$ from it, mimicking the ``Dirac revival'' inferred from the same behavior in MATBG~\cite{Zondiner2020}. Here, however, the itinerant electrons and the local moments both come from the orbital $\hf$. In the spectral function a flat Hubbard band sits at the Fermi level between two integers and resets at each one~\cite{Wong2020}. With larger $s^2$ the maxima of $\kappa^{-1}$ move away from the integers, and a Kondo resonance stays pinned at the Fermi level at every filling, which signals a heavy Fermi liquid [Fig.~\ref{fig:cascade}(c,f)]. This suggests an insulator-to-metal transition tuned by $s^2$ at non-zero integer filling.

\emph{Mott transition.}---We illustrate the transition at $\nu=-2$. The charge gap closes near the transition, which we place at $s^2_c\simeq0.025$ [Fig.~\ref{fig:mott}(a,b)], and the Hubbard-I and Hubbard-III approximations fail before the transition. The weight at the Fermi level at the zone corner, $Q_K=-\beta G_c(K,\beta/2)/\pi$, falls on cooling below $s^2_c$ and grows above it [Fig.~\ref{fig:mott}(b)]. Below $s^2_c$ the electron spectrum has a clear gap. Above it the moments are Kondo screened, and a heavy band crosses the Fermi level [Fig.~\ref{fig:mott}(c,d)]. The gap closes as the band of $F$ moves up and touches the Fermi level, where it screens the local moments, so that the metal is a heavy Fermi liquid below $T_K$ (Appendix~\ref{app:mott}). At $\nu=-1$ and $-3$ a similar transition lies at $s^2_c\simeq0.033$ and $0.027$ (Appendix~\ref{app:results}).

\begin{figure}[t]
 \includegraphics[width=\columnwidth]{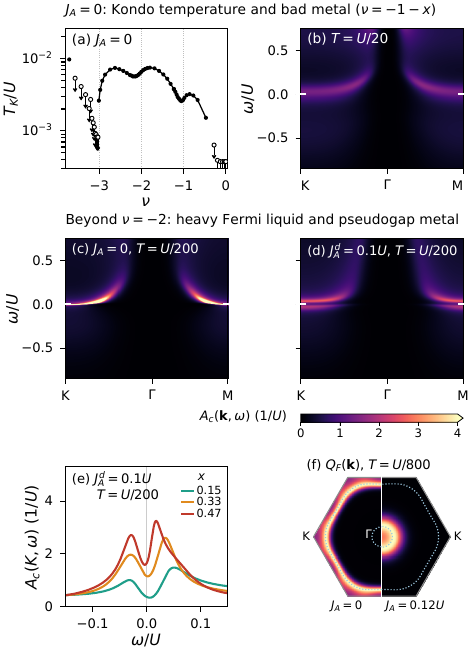}
 \caption{The doped metal ($\Nf=8$, $s^2=0.05$). (a) Kondo temperature across the filling range without the anti-Hund's coupling. Filled symbols mark a saturated spin susceptibility, and open symbols with arrows are upper limits of $T_K$. (b) The bad metal: $A_c(\bk,\omega)$ at $x\simeq0.2$ beyond $\nu=-1$ without the anti-Hund's coupling, at $T=U/20\simeq13\,T_K$. The color bar under (d) is shared by (b--d) and saturated above $4/U$. (c,d) $A_c(\bk,\omega)$ beyond $\nu=-2$ at $T=U/200$, for the heavy Fermi liquid without the anti-Hund's coupling ($x=0.34$, $T\simeq0.8\,T_K$) and for the pseudogap metal with $J_A^d=0.1U$ of Eq.~\eqref{eq:had} ($x=0.33$). (e) Pseudogap at the Brillouin zone corner, $A_c(K,\omega)$, with $J_A^d=0.1U$ at $T=U/200$ for $x=0.15$, $0.33$ and $0.47$. Panels (b--e) use maximum entropy method for analytical continuation. (f) Fermi surface at $U/800$ without analytical continuation. It shows the weight of $F$ at the Fermi level, $Q_F(\bk)=-\beta G_F(\bk,\beta/2)/\pi$, at $x\simeq0.2$ beyond $\nu=-2$ without (left) and with $J_A=0.12U$ of Eq.~\eqref{eq:ha} (right), each  normalized to its maximum. The dotted lines are the contours that enclose $x$ and $6+x$ holes.}
 \label{fig:metal}
\end{figure}

\emph{Heavy Fermi liquid and bad metal.}---Beyond $s^2_c$ the moments are screened at every filling [Fig.~\ref{fig:metal}(a)]. Below a small Kondo temperature $T_K$ the state is always a heavy Fermi liquid (HFL), with a large Fermi surface that satisfies the Luttinger theorem [Fig.~\ref{fig:metal}(c,f)]. We locate the Fermi surface without continuation, as the contour on which the weight at the Fermi level, $Q_\beta(\bk)=-\beta G(\bk,\beta/2)/\pi$ of the electron or of $F$, is largest. Above $T_K$ the local moments are not fully screened, and there is no Fermi surface because the incipient heavy band lies above the Fermi level [Fig.~\ref{fig:metal}(b)]. The crossover from the bad metal to the HFL on cooling is captured in a unified framework by the ancilla theory with $c$, $F$ and $F'$ [Fig.~\ref{fig:intro}(b)], in which $\Sigma_F$ of Eq.~\eqref{eq:dysonF} becomes
\begin{equation}
 \Sigma_F(z)\simeq\frac{B^2}{z-\epsilon_{F'}+i\Gamma_{F'}}+\Sigma_{\rm inc}(z).
 \label{eq:chain}
\end{equation}
The electron hybridizes with $F$, and $F$ with the spinon $F'$ through $B(F^\dagger_iF'_i+{\rm H.c.})$, the standard mean-field description of the Kondo-screened phase~\cite{Coleman2015}. We fit these parameters to the DMFT self-energy. $\Sigma_{\rm inc}$ is mainly the Hubbard-III scattering rate of $F$, and the fits place the spinon level below the Fermi level, $\epsilon_{F'}<0$. At $T\ll T_K$, where $\Gamma_{F'}\to0$ and $\Sigma_{\rm inc}$ vanishes at the Fermi level, there emerges the heavy band, mostly the combination of $c$ and $F'$, which crosses the Fermi level on the large Fermi surface. Above $T_K$ the spinon $F'$ acquires a scattering rate $\Gamma_{F'}\propto T$, which sets the linear-in-$T$ scattering rate of the electron, $-\Impart\Sig(0)\simeq(\Phi^2/B^2)\Gamma_{F'}\approx3.5$--$3.7\,k_BT$ at $s^2=0.05$. $F$ also decays fast at the Fermi level. The bad metal has the heavy band above the Fermi level, with a width linear in temperature, about $3\,k_BT$ at half maximum. It remains to be seen whether the linear-in-$T$ resistivity observed in MATBG~\cite{Polshyn2019,Cao2020strange,Jaoui2022} emerges above $T_K$ in the bad metal.

\emph{Pseudogap metal.}---At $\nu=-2-x$ an anti-Hund's coupling changes the physics qualitatively. In MATBG it favors inter-valley spin singlets~\cite{ZhaoZhang2025}, and we model it by a simple form,
\begin{equation}
 \begin{gathered}
 H_A^d=-J_A^d\sum_{i,o=\pm}\Pi_{io}^\dagger\Pi_{io},\\
 \Pi_o^\dagger=\tfrac{1}{\sqrt2}\big(\hf^\dagger_{oK\uparrow}\hf^\dagger_{oK'\downarrow}-\hf^\dagger_{oK\downarrow}\hf^\dagger_{oK'\uparrow}\big),
 \end{gathered}
 \label{eq:had}
\end{equation}
where $o=\pm$ is the orbital of angular momentum $L_z=\pm1$, so that the term favors intra-orbital inter-valley spin singlets with $L_z=\pm2$. We solve the impurity by CT-HYB in its matrix form~\cite{Seth2016} (Appendix~\ref{app:pg}). Once the coupling exceeds a few times $T_K$, the moments form local singlets that disrupt the Kondo screening, and a pseudogap of size $\sim J_A^d$ opens in the heavy band at the Fermi level [Fig.~\ref{fig:metal}(d,e)]. Another term, $J_A$ of Eq.~\eqref{eq:ha}, which favors an inter-orbital singlet, gives very similar pseudogap physics, and we use it for the fits below and for Fig.~\ref{fig:metal}(f). To capture the crossover from the bad metal to the pseudogap metal, we introduce a second pair of ancilla fermions, $\psi$ and $\psi'$, besides $F$ and $F'$ [Fig.~\ref{fig:intro}(b)]. The fermions $\psi'$ are spinons bound into a product of local singlets, and the spinon $F'$ of Eq.~\eqref{eq:chain} hybridizes with $\psi$ through $\Phi_s(F'^\dagger_i\psi_i+{\rm H.c.})$, which opens the pseudogap, $\Delta_{\rm PG}\sim\Phi_s$. The self-energy of $F$ becomes
\begin{equation}
 \Sigma_F(z)=\cfrac{B^2}{z-\epsilon_{F'}-\cfrac{\Phi_s^2}{z-\epsilon_\psi+i\Gamma_\psi}}+\Sigma_{\rm inc}(z),
 \label{eq:pgchain}
\end{equation}
which fits the DMFT self-energy to a few percent. The pseudogap opens below the temperature $T^*$ of Fig.~\ref{fig:intro}(c), of order $J_A^d$. Between $T^*$ and a much smaller coherence temperature, $T_{\rm coh}\sim U/800$, the band of $\psi$ lies just below the Fermi level, and again there is no Fermi surface. Below $T_{\rm coh}$ the top of this band, at $\Gamma$, rises above the Fermi level, and quasiparticles carried mainly by $\psi$ form a small Fermi surface of about $x$ holes [Fig.~\ref{fig:metal}(f)], which gives the second Fermi liquid (sFL) phase of Ref.~\cite{ZhaoZhang2025}. We identify $\psi$ as a spin polaron, an electron bound to the spin of the other valley. When we allow inter-valley coherent (IVC) order in DMFT, it develops only below $T^*$, so symmetry breaking is not essential for the formation of the pseudogap.

\emph{Summary.}---A model with only the Hubbard $U$ on coherent orbitals captures many essential features of MATBG. It reproduces the cascade and the sawtooth of the inverse compressibility without remote bands or a Dirac revival. Instead, a heavy-fermion picture emerges, in which the itinerant electrons and the local moments come from the same orbital $\hf$, and the Kondo coupling between them is controlled by the overlap $s^2$ of the coherent orbitals, set by how strongly the Berry curvature is concentrated. At non-zero integer filling the Mott insulator turns into a metal through Kondo screening. At $\nu=-2-x$ the normal state is a bad metal at high temperature, which crosses over to a pseudogap metal below $T^*$, and we expect the pseudogap metal to be the normal state of the superconductor~\cite{Cao2018b,ZhaoZhang2025}. It becomes a Fermi liquid with a small Fermi surface only below $T_{\rm coh}\sim U/800$, about $0.43$~K for $U\approx30$~meV, which appears to lie below the critical temperature of the superconductor. Thus, at $T_c$, a metal without a Fermi surface becomes superconducting, with coherent quasiparticles developed only at lower temperature. This is a situation beyond the conventional picture of superconductivity that needs further study. More broadly, we develop a framework to study Mott physics in topological bands, which we expect to be relevant for other systems such as twisted WSe$_2$~\cite{Wu2019tmd,Wang2020tmd,Ghiotto2021,Xia2025,Guo2025}, where the kinetic energy is not negligible and adds to the quantum geometry in driving the Mott transition.

\emph{Acknowledgments.}---I used Claude Opus 5.5 to implement the DMFT calculations using the TRIQS library~\cite{Parcollet2015}. The ideas of single-site DMFT on coherent orbitals and its interpretation in the ancilla framework are my own. This work was supported by the National Science Foundation under Grant No.~DMR-2237031.

\bibliography{references}

\clearpage
\appendix
\setcounter{secnumdepth}{2}
\onecolumngrid

\section{Coherent orbitals of a flat Chern band}
\label{app:coherent}
This appendix defines the coherent orbitals and gives two realizations, the flat bands of MATBG and a lowest Landau level in a modulated magnetic field.

\subsection{Projected orbitals and the distribution of their weights}

\emph{Projected orbitals.}---Each of the $\Nf$ flavors has a flat band of Chern number $C=\pm1$ with Bloch states $|\psi_{\bk}\rangle$. We place the same localized orbital $|\tau_i\rangle$ on every site $\bm R_i$. Its Bloch sum $|\tau_{\bk}\rangle=N_M^{-1/2}\sum_ie^{i\bk\cdot\bm R_i}|\tau_i\rangle$ over the $N_M$ unit cells overlaps only with the Bloch state of the same momentum, with the form factor $\mu_f(\bk)=\langle\tau_{\bk}|\psi_{\bk}\rangle$. The weight $w_f(\bk)=|\mu_f(\bk)|^2$ has the zone average $W_f$. Projecting $\tau_i$ onto the band and normalizing gives the coherent orbital
\begin{equation}
 \hf_{i}=\frac1{\sqrt{N_M}}\sum_{\bk}e^{i\bk\cdot\bm R_i}\,\frac{\mu_f(\bk)}{\sqrt{W_f}}\,c_{\bk},\qquad
 S_{\bk}=\frac{w_f(\bk)}{W_f},
 \label{eq:projection}
\end{equation}
where $c_{\bk}$ annihilates an electron in $|\psi_{\bk}\rangle$. The orbital is normalized, $\avg{S_{\bk}}=1$. It overlaps with the orbitals on other sites, $\{\hf_{i\alpha},\hf^\dagger_{j\beta}\}=\delta_{\alpha\beta}S_{ij}$ with $S_{ij}=N_M^{-1}\sum_{\bk}e^{i\bk\cdot(\bm R_i-\bm R_j)}S_{\bk}$.

\emph{The overlap is forced by the topology.}---In a band with $|C|=1$ the form factor of any localized orbital vanishes at some momentum $\bk_0$, around which its phase winds once~\cite{Perelomov1971,Bacry1975}. Otherwise $\mu_f/|\mu_f|$ would be a smooth periodic gauge and give exponentially localized Wannier functions, which a Chern band does not admit~\cite{Brouder2007}. The coherent orbitals are therefore overcomplete by exactly one state. Their overlap $\eta_S^2=\sum_{j\ne i}|S_{ij}|^2=\avg{S_{\bk}^2}-1$ is always positive, but it can be made arbitrarily small if the Berry curvature is concentrated.

\begin{figure*}[t]
 \includegraphics[width=\textwidth]{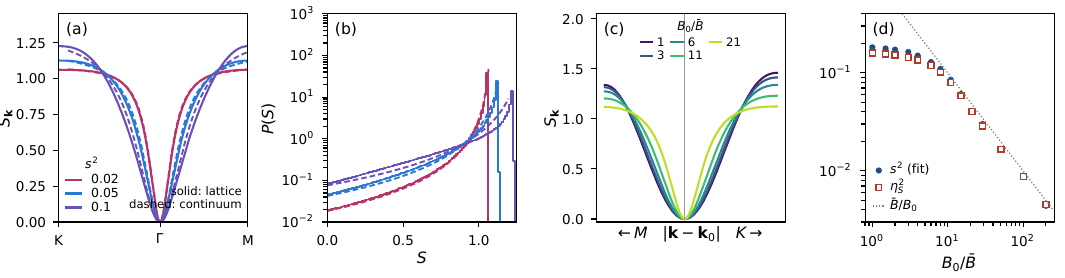}
 \caption{Weight of the coherent orbital. (a) $S_{\bk}$ along $K$--$\Gamma$--$M$ for the lattice texture of Eq.~\eqref{eq:wlat} (solid) and the continuum Lorentzian $w_f=k^2/(k^2+k_*^2)$ with the same $s^2$ (dashed). (b) Their distributions $P(S)$, the histogram of the lattice texture and Eq.~\eqref{eq:PSlor}. (c) $S_{\bk}$ of the lowest Landau level in a modulated field, Eq.~\eqref{eq:Sfluct}, from its zero $\bk_0$ toward $M$ and $K$, for the enhancements $B_0/\bar B$ of the field on the sites given in the legend. $B_0=\bar B$ is the uniform field, Eq.~\eqref{eq:Slll}. (d) $s^2$ from a Lorentzian fit of $S_{\bk}$ around its zero, and $\eta_S^2$, versus $B_0/\bar B$. The dotted line is $\bar B/B_0$.}
 \label{fig:texture}
\end{figure*}

\emph{Distribution of the orbital weights.}---DMFT sees the band only through the distribution of the orbital weights,
\begin{equation}
 P(S)=\frac1{N_M}\sum_{\bk}\delta(S-S_{\bk}),
 \label{eq:PS}
\end{equation}
with $\int\dd S\,P(S)=\int\dd S\,S\,P(S)=1$ and $\int\dd S\,(S-1)^2P(S)=\eta_S^2$. Near $\bk_0=0$ the weight is $w_f(\bk)\simeq k^2/(k^2+k_*^2)$. Then $s^2=\pi k_*^2/\Omega_{\rm BZ}$ is the fraction of the zone in which $w_f<1/2$. This Lorentzian form gives
\begin{equation}
 P(S)=\frac{W_f\,s^2}{(1-W_fS)^2},\qquad 0\le S\le S_{\max}=\frac1{W_f(1+s^2)},\qquad W_f=1-s^2\ln\Big(1+\frac1{s^2}\Big).
 \label{eq:PSlor}
\end{equation}
To leading order in $s^2$ it is $P(S)=s^2/(1-S)^2$ on $0<S<1$. The mixed states thus carry the uniform weight $P(S)(1-S)^2=s^2$. The overlap is $\eta_S^2\simeq s^2$. In DMFT we use the lattice version of this texture on the triangular lattice of unit spacing,
\begin{equation}
 \begin{gathered}
  w_f(\bk)=\frac12\Big(1-\frac{m_{\bk}}{\sqrt{m_{\bk}^2+4|g_{\bk}|^2/k_*^2}}\Big),\\
  g_{\bk}=\frac23\sum_{j=0}^{2}e^{-i\theta_j}\sin(\bk\cdot\bm\delta_j),\qquad m_{\bk}=1-\frac{4}{3k_*^2}\sum_{j=0}^{2}\big[1-\cos(\bk\cdot\bm\delta_j)\big],
 \end{gathered}
 \label{eq:wlat}
\end{equation}
with $\bm\delta_j=(\cos\theta_j,\sin\theta_j)$ and $\theta_j=2\pi j/3$. It reduces to $k^2/(k^2+k_*^2)$ near $\bk=0$ and is close to the continuum form [Fig.~\ref{fig:texture}(a,b)].

\subsection{Twisted bilayer graphene}

\emph{Flavors.}---In MATBG the flat bands of each spin and valley can be rotated into two sectors of Chern number $\pm1$~\cite{Tarnopolsky2019,Ledwith2020,Wang2021}, which gives $\Nf=8$ flavors. $C_2\mathcal T$ relates the two Chern sectors of a valley, and time reversal relates the two valleys. Both keep the orbital weights up to $\bk\to-\bk$, so all eight flavors share the same $P(S)$.

\emph{Coherent orbital from the continuum model.}---The $f$ orbital of the topological heavy-fermion model~\cite{Song2022}, projected onto the flat bands, can serve as $\hf$. The flat bands alone suffice, however~\cite{ZhaoAncilla2025}. In each valley we rotate the two flat Bistritzer--MacDonald bands at the first magic angle into Chern bands, the eigenvectors of the projected sublattice operator $\sigma_z$. They carry $C=\pm1$ and are polarized on opposite sublattices. For the $C=+1$ band the orbital $\tau$ is a pair of narrow Gaussian wave packets on an AA site, one in each layer, on the sublattice of that band. Its weight is $w_f(\bk)=\sum_{l}|\langle\tau_{l\bk}|\psi_{\bk}\rangle|^2$, with $|\tau_{l\bk}\rangle$ the Bloch sum of the packet in layer $l$. This weight vanishes at $\Gamma$, where the phase of the form factor $\mu_f$ winds. Here $s^2$ is the fraction of the zone in which $w_f$ lies below half its maximum. Figure~\ref{fig:intro}(a) shows the resulting $S_{\bk}$, whose $P(S)$ is close to the Lorentzian one with the same $s^2$.

\subsection{Lowest Landau level in a modulated magnetic field}

The physics of this paper should follow from a Hubbard interaction on the coherent orbitals of any flat Chern band, not only MATBG. Here $s^2$ is tuned continuously while the band stays exactly flat. Consider an electron with $g=2$ in a periodic field $B(\bm r)\hat z$, with one flux quantum per unit cell of a triangular lattice of sites $\bm R_i$. Its Pauli Hamiltonian is
\begin{equation}
 H=\frac{[\bm\sigma\cdot(\bm p+e\bm A)]^2}{2m}=\frac{(\bm p+e\bm A)^2}{2m}+\frac{e\hbar}{2m}B(\bm r)\,\sigma_z .
 \label{eq:pauli}
\end{equation}
For one spin component the Zeeman energy cancels the zero-point energy of the cyclotron motion. The zero-energy states then form an exactly flat band with $C=1$ for any $B(\bm r)$ of one sign~\cite{AharonovCasher1979}. For a uniform field it is the lowest Landau level.

\emph{Uniform field.}---For $B=\bar B$ the unit cell has the area $2\pi\ell^2$, with $\ell=(\hbar/e\bar B)^{1/2}$. Projecting the delta function at $\bm R_i$ onto the level gives the coherent state~\cite{Perelomov1971,Bacry1975} $\varphi_i(\bm r)\propto\exp[-|\bm r-\bm R_i|^2/4\ell^2+i(\bm R_i\times\bm r)_z/2\ell^2]$ in the symmetric gauge. These states overlap as $S_{ij}=\epsilon_{\bm R_i-\bm R_j}\,e^{-|\bm R_i-\bm R_j|^2/4\ell^2}$, where $\epsilon_{\bm R}=(-1)^{mn}$ is the sign of the magnetic translation by $\bm R=m\bm a_1+n\bm a_2$. Hence
\begin{equation}
 S^{(0)}_{\bk}=\sum_{\bm R}\epsilon_{\bm R}\,e^{-R^2/4\ell^2}\cos(\bk\cdot\bm R) .
 \label{eq:Slll}
\end{equation}
It vanishes only at $\bk_0=(\bm b_1+\bm b_2)/2$ in this gauge. The overlap is $\eta_S^2=\sum_{\bm R\neq0}e^{-R^2/2\ell^2}$. This value and $s^2$, from a Lorentzian fit around $\bk_0$, are the leftmost points of Fig.~\ref{fig:texture}(d).

\emph{Modulated field.}---Now let the field vary at fixed flux per cell, $B(\bm r)=\bar B+\delta B(\bm r)$ with $\delta B=(\Delta B/3)\sum_j\cos(\bm G_j\cdot\bm r)$ over the three shortest reciprocal-lattice vectors. The field peaks on the sites at $B_0=\bar B+\Delta B$. Following Aharonov and Casher~\cite{AharonovCasher1979}, we write $\bm A=(\hbar/e)\,\hat z\times\nabla\varphi$ with $\nabla^2\varphi=eB/\hbar$. The zero modes are then $\psi=f(z)\,e^{-\varphi}$, with $f$ analytic in $z=x+iy$. The uniform part of $\varphi$, $|\bm r|^2/4\ell^2$, gives the lowest Landau level. Its periodic part $\delta\varphi$ multiplies every zero mode by the same positive function,
\begin{equation}
 \psi_{\bk}(\bm r)=\psi^{(0)}_{\bk}(\bm r)\,e^{-\delta\varphi(\bm r)},\qquad
 \nabla^2\delta\varphi=\frac{e\,\delta B}{\hbar},
 \label{eq:AC}
\end{equation}
with $\psi^{(0)}_{\bk}$ the Bloch states of the uniform field. The analytic factors, and with them the zeros and the Chern number, are unchanged. Only the density is reweighted, by $e^{-2\delta\varphi}$, which peaks on the sites. Because only the norm of the zero modes changes, the projected delta function at $\bm R_i$ has the weight
\begin{equation}
 S_{\bk}\propto\frac{S^{(0)}_{\bk}}{N_{\bk}},\qquad
 N_{\bk}=\int_{\rm cell}\dd^2r\,|\psi^{(0)}_{\bk}(\bm r)|^2e^{-2\delta\varphi(\bm r)} .
 \label{eq:Sfluct}
\end{equation}
Near a field maximum the orbital is a Gaussian of width $\sqrt2\ell_0$, with $\ell_0=\ell(\bar B/B_0)^{1/2}$ the local magnetic length. Hence $k_*=\sqrt2\ell_0/\ell^2$ and $s^2\simeq\bar B/B_0$. Concentrating the field shrinks the orbital and sharpens its weight around the zero [Fig.~\ref{fig:texture}(c)]. For $B_0\gg\bar B$ both $s^2$ and $\eta_S^2$ fall as $\bar B/B_0$ [Fig.~\ref{fig:texture}(d)], while the band stays exactly flat.

\emph{Quantum geometry.}---Every band of this family is ideal, $\operatorname{tr}g=\Omega$. Its integrated metric $\int\operatorname{tr}g\,\dd^2k=2\pi$ is therefore pinned and does not measure the overlap. What changes is the distribution of the Berry curvature. For the Lorentzian texture the weight within radius $k$ of the zero is the Berry flux it encloses, $w_f(k)=(2\pi)^{-1}|\int_{|\bm q|<k}\Omega\,\dd^2q|$. The participation ratio of the curvature is then $\avg{\Omega}^2/\avg{\Omega^2}=3s^2$, so $\eta_S^2\simeq\avg{\Omega}^2/3\avg{\Omega^2}$. The coherent orbitals thus overlap little when the Berry curvature is concentrated.
\section{DMFT on coherent orbitals}
\label{app:dmft}

\subsection{Impurity problem and self-consistency}

\emph{Impurity model.}---We consider the Hubbard model on the coherent orbitals,
\begin{equation}
 H=-\mu\sum_{\bk}c^\dagger_{\bk}c_{\bk}
 +\frac U2\sum_i\Big(\hat n_{fi}-\frac{\Nf}2\Big)^2,
 \label{eq:model}
\end{equation}
with $\hat n_{fi}=\sum_\alpha\hf^\dagger_{i\alpha}\hf_{i\alpha}$. The band is exactly flat, so the electrons have no kinetic energy of their own. The chemical potential couples to the band filling per site, $n_c$, with $\nu=n_c-\Nf/2$. Energies are in units of $U$. DMFT~\cite{Georges1996} replaces the full model by an impurity model on a single coherent orbital in a self-consistent bath. The action of the impurity site is
\begin{equation}
 \begin{aligned}
 \mathcal S_{\rm imp}={}&-\int_0^\beta\!\!\dd\tau\,\dd\tau'\,\bar f_\alpha(\tau)\,\mathcal G_0^{-1}(\tau-\tau')\,f_\alpha(\tau')\\
 &+\frac U2\int_0^\beta\!\!\dd\tau\,\Big(n_f(\tau)-\frac{\Nf}2\Big)^2,
 \end{aligned}
 \label{eq:Simp}
\end{equation}
with flavors summed and $\mathcal G_0^{-1}(i\omega_n)=i\omega_n+\mu-\Delta_f(i\omega_n)$. The hybridization function $\Delta_f(\tau)$ captures the effects of the other sites in the system. The impurity propagator, $G_{\rm imp}(\tau)=-\langle f_\alpha(\tau)\bar f_\alpha(0)\rangle$, defines the impurity self-energy $\Sig=\mathcal G_0^{-1}-G_{\rm imp}^{-1}$. The DMFT approximation says that the lattice self-energy is from this impurity site: $\Sig(z)\sum_i\hf^\dagger_{i\alpha}\hf_{i\alpha}$. By Eq.~\eqref{eq:projection}, $\sum_i\hf^\dagger_{i\alpha}\hf_{i\alpha}=\sum_{\bk}S_{\bk}\,c^\dagger_{\bk\alpha}c_{\bk\alpha}$, so DMFT actually gives a self-energy with momentum dependence in the band basis. With $\zeta=z+\mu$ the band electron and the local orbital have the propagators
\begin{equation}
 \begin{gathered}
 G_c(\bk,z)=\frac1{\zeta-S_{\bk}\Sig(z)},\\
 \Gloc_f(z)=\frac1{N_M}\sum_{\bk}S_{\bk}\,G_c(\bk,z)=\int\dd S\,\frac{P(S)\,S}{\zeta-S\Sig(z)} .
 \end{gathered}
 \label{eq:Gc}
\end{equation}
Each Bloch state feels the interaction in proportion to its orbital weight, and the state at the zero does not feel it at all. The impurity must reproduce the local propagator of the orbital in the lattice, $G_{\rm imp}=\Gloc_f$, which gives the self-consistency equation
\begin{equation}
 \Delta_f(z)=\zeta-\Sig(z)-\frac1{\Gloc_f(z)} .
 \label{eq:weiss}
\end{equation}
Equations~\eqref{eq:Simp}--\eqref{eq:weiss} are solved by iteration: starting from the self-energy of the isolated atom, one computes $\Gloc_f$ and $\Delta_f$, solves the impurity problem for a new $\Sig$, mixes it with the previous one, and repeats until $\Delta_f$ no longer changes within the statistical errors, typically after 15--30 iterations. 

\emph{Relation to DFT+DMFT.}---The projection $\Gloc_f=\avg{S_{\bk}G_c}$ and the embedding of $\Sig$ as $S_{\bk}\Sig$ have the form used in DFT+DMFT with projected local orbitals~\cite{Haule2010}. There the correlated orbitals lie almost entirely within a wide window of Bloch bands, and their small non-orthonormality, which comes from the truncated window and the overcomplete basis, is removed or corrected by orthonormalizing or renormalizing the projector. Here the window is a single Chern band, which contains no orthonormal localized orbitals. The orbitals are therefore normalized only on site, $\avg{S_{\bk}}=1$, and their overlap is enforced by the topology. This overlap is the only source for the bath $\Delta_f(z)$ for the flat band because we ignore the dispersion in this work.

\subsection{Where the bath comes from}

\emph{Exact form.}---Equation~\eqref{eq:weiss} gives $\Delta_f$ as the difference of two terms that nearly cancel. Carrying out the subtraction under the integral of Eq.~\eqref{eq:Gc} shows what the bath is made of. Write Eq.~\eqref{eq:weiss} as $\Delta_f=[(\zeta-\Sig)\Gloc_f-1]/\Gloc_f$ and use $\int\dd S\,P(S)=1$ to bring the numerator under that integral,
\begin{equation}
 (\zeta-\Sig)\Gloc_f-1=\int\dd S\,P(S)\,\frac{(\zeta-\Sig)S-(\zeta-S\Sig)}{\zeta-S\Sig}=\zeta\int\dd S\,P(S)\,\frac{S-1}{\zeta-S\Sig} .
\end{equation}
Since also $\int\dd S\,P(S)\,S=1$, so that $\int\dd S\,P(S)\,(S-1)=0$, the last integral does not change when $(S-1)/(\zeta-\Sig)$ is subtracted from its integrand. With
\begin{equation}
 \frac{S-1}{\zeta-S\Sig}-\frac{S-1}{\zeta-\Sig}=\frac{\Sig\,(1-S)^2}{(\zeta-\Sig)(\zeta-S\Sig)}
\end{equation}
the Weiss field becomes
\begin{equation}
 \Delta_f(z)=\frac{\zeta\,\Sig}{(\zeta-\Sig)\,\Gloc_f}\int\dd S\,P(S)\,\frac{(1-S)^2}{\zeta-S\Sig} .
 \label{eq:Deltaexact}
\end{equation}

\emph{What the bath is.}---Equation~\eqref{eq:Deltaexact} is an integral over the orbital weights with the weight $P(S)(1-S)^2$, which vanishes in the flat region $S\simeq1$ and integrates to $\eta_S^2\simeq s^2$: the orbital exchanges electrons with the rest of the lattice only through the mixed Bloch states, $0<S<1$, and a lattice of orthonormal orbitals, $S_{\bk}\equiv1$, has no bath. At large $z$, $\Gloc_f\to1/\zeta$ and $\Delta_f\to\Delta_\infty=\eta_S^2\Sigma_\infty$, with the Hartree constant $\Sigma_\infty=U(\Nf-1)(\langle n_\alpha\rangle-\tfrac12)$. The impurity level is $\mu_{\rm imp}=\mu-\Delta_\infty$.

\section{Expansion in \texorpdfstring{$s^2$}{s2}: Hubbard-I, Hubbard-III and the Kondo coupling}
\label{app:expansion}
The bath of Appendix~\ref{app:dmft} is of order $s^2$, so we can formulate a perturbation theory by truncating the DMFT self-consistency equation. At zeroth order we reproduce the Hubbard-I approximation and at order $s^2$ the Hubbard-III approximation. Interestingly, from the analytical derivation here we arrive at the ancilla framework, in which the electron $c_{\bk}$ couples to a pair of auxiliary fermions $(F,F')$. Hubbard-I can be viewed as the hybridization between $c$ and $F$, and Hubbard-III gives a scattering rate to the $F$ fermion. However, both approximations ignore the $F'$ fermion, which is the spinon of the local moments. In our formulation, we find that the bath of DMFT is dominated by the $F$ fermion, $\rho_\Delta(\omega)=\Phi^2\rho_F(\omega)$ [Eq.~\eqref{eq:rhoF}], and that there is a Kondo coupling $J_K\sim U$ between the $F$ fermion and the local moments represented by $F'$. Hubbard-III can be easily derived from Fermi's golden rule for this Kondo coupling. But the Kondo coupling is marginally relevant and flows to strong coupling at the Kondo temperature $T_K$. Therefore we expect the perturbative expansion in $s^2$ to break down at larger $s^2$ because of Kondo screening. The fully self-consistent DMFT calculation beyond this appendix finds an insulator-to-metal transition at nonzero integer filling around $s^2\simeq0.025$--$0.033$. To describe the corresponding heavy Fermi liquid (HFL), we need to keep both $(F,F')$ in the effective theory.

The prefactor of Eq.~\eqref{eq:Deltaexact} is $\zeta\Sig+O(s^2)$, because $\Gloc_f=1/(\zeta-\Sig)+O(s^2)$. Hence
\begin{equation}
 \begin{gathered}
 \Delta_f(z)=0+s^2\Delta_f^{(1)}(z)+O(s^4),\\
 \Delta_f^{(1)}=\zeta\Sig\int\dd S\,\frac{P(S)(1-S)^2/s^2}{\zeta-S\Sig},
 \end{gathered}
 \label{eq:Dexpand}
\end{equation}
where $P(S)(1-S)^2/s^2$ does not depend on $s^2$ for the Lorentzian form of $S_{\bk}$ at this order, Eq.~\eqref{eq:PSlor}. The impurity self-energy can then be expanded in terms of $s^2$.

\subsection{Zeroth order: Hubbard-I and the orthogonal fermion}
\label{app:hubbard1}

\emph{Hubbard-I.}---At zeroth order we use $\Delta_f(z)=0$ and then every site is an isolated SU($\Nf$) atom. Let it hold $n_f$ electrons, an integer fixed by the chemical potential through $U(n_f-\Nf/2-\tfrac12)<\mu<U(n_f-\Nf/2+\tfrac12)$, and let $q=n_f/\Nf$. The atomic self-energy is
\begin{equation}
 \begin{gathered}
 \Sigma_{\rm at}(z)=\Sigma_\infty+\frac{\Phi^2}{z-\epsilon_F},\\
 \Sigma_\infty=U(\Nf-1)(q-\tfrac12),\qquad \Phi^2=U^2q(1-q),
 \end{gathered}
 \label{eq:sigat}
\end{equation}
with $\epsilon_F=qE_+-(1-q)E_-$.  With the atomic energies $E_{\rm at}(n)=\frac U2(n-\frac{\Nf}2)^2-\mu n$, we have 
\begin{equation}
 \begin{aligned}
 E_+&=E_{\rm at}(n_f+1)-E_{\rm at}(n_f)=U\big(n_f-\tfrac{\Nf}2+\tfrac12\big)-\mu,\\
 E_-&=E_{\rm at}(n_f-1)-E_{\rm at}(n_f)=U\big(\tfrac{\Nf}2+\tfrac12-n_f\big)+\mu,
 \end{aligned}
 \label{eq:Epm}
\end{equation}
that is, $E_\pm=U/2\pm[U(n_f-\Nf/2)-\mu]$, with $E_++E_-=U$. In DMFT $\mu$ is replaced by the level of the impurity, $\mu_{\rm imp}=\mu-\Delta_\infty$. 

The pole of $\Sigma_{\rm at}$ may be viewed as the hybridization of the physical electron $\hf$ with another fermion $F$. This may be seen from the atomic Hamiltonian $H_{\rm at}=\tfrac U2(n-\Nf/2)^2-\mu_{\rm imp}n$, which gives exactly $[\hf_\alpha,H_{\rm at}]=[U(N'_\alpha-\tfrac{\Nf-1}2)-\mu_{\rm imp}]\hf_\alpha$, with $N'_\alpha=\sum_{\beta\neq\alpha}n_\beta$. The part of it beyond the mean charge defines a new fermion, $F_\alpha=(U/\Phi)\,\delta N'_\alpha\,\hf_\alpha$, where $\delta N'_\alpha=N'_\alpha-\langle N'_\alpha\rangle$ is the charge of the other flavors measured from its mean. With $\Phi^2=U^2\langle\delta N_\alpha'^2\rangle$ the prefactor normalizes it, $\langle\{F_\alpha,F^\dagger_\alpha\}\rangle=1$. Then, exactly,
\begin{equation}
 [\hf_\alpha,H_{\rm at}]=(\Sigma_\infty-\mu_{\rm imp})\hf_\alpha+\Phi\,F_\alpha,\qquad
 [F_\alpha,H_{\rm at}]=\frac U\Phi\big[U\delta N_\alpha'^2+(\Sigma_\infty-\mu_{\rm imp})\,\delta N'_\alpha\big]\hf_\alpha,
 \label{eq:eomexact}
\end{equation}
with $\Sigma_\infty=U(\langle N'_\alpha\rangle-\tfrac{\Nf-1}2)$. The first relation closes on $\hf$ and $F$; the second does so only in a subspace. At zeroth order the atom holds $n_f$ electrons, and its Green functions involve only its states with $n_f-1$, $n_f$ and $n_f+1$ electrons. Let
\begin{equation}
 P_G=P_{n_f-1}+P_{n_f}+P_{n_f+1}
 \label{eq:PG}
\end{equation}
project on them, with $P_n$ the projector on the states with $n$ electrons. Inside $P_G$ an operator that removes the flavor $\alpha$ acts on states with $n_f$ or $n_f+1$ electrons, on which the charge of the other flavors is $N'_\alpha=n_f-1$ or $n_f$. The variable $N'_\alpha-(n_f-1)$ is thus $0$ or $1$, with the mean $1-q$, since $\langle N'_\alpha\rangle=(\Nf-1)q=n_f-q$. For such a variable $\delta N_\alpha'^2=q(1-q)+(2q-1)\,\delta N'_\alpha$, and the bracket of Eq.~\eqref{eq:eomexact} becomes $Uq(1-q)+[U(2q-1)+\Sigma_\infty-\mu_{\rm imp}]\,\delta N'_\alpha$. With $\Phi^2=U^2q(1-q)$ in the multiplet, the projected operators $\hf^G_\alpha=P_G\hf_\alpha P_G$ and $F^G_\alpha=P_GF_\alpha P_G$ therefore obey
\begin{equation}
 [\hf^G_\alpha,H_{\rm at}]=(\Sigma_\infty-\mu_{\rm imp})\hf^G_\alpha+\Phi\,F^G_\alpha,\qquad
 [F^G_\alpha,H_{\rm at}]=\Phi\hf^G_\alpha+\epsilon_F F^G_\alpha,\qquad
 \epsilon_F=\Sigma_\infty-\mu_{\rm imp}+U(2q-1),
 \label{eq:eomat}
\end{equation}
 The propagator of $\hf^G$ is the same as that of $\hf$ without projection for the atomic Mott insulator. Outside the $P_G$ subspace, on states with $n_f\pm2$ electrons, the equations of motion do not close, and a theory with $(\hf^G,F^G)$ alone is not sufficient.

\begin{figure}[t]
 \includegraphics[width=0.45\linewidth]{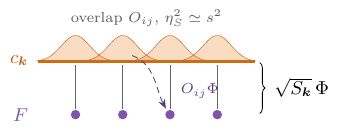}
 \caption{Mean-field theory of $c$ and $F$, Eq.~\eqref{eq:hmf}, in the Hubbard-I approximation. The band electrons $c_{\bk}$ (lobes, the overlapping coherent orbitals) hybridize with the orthogonal fermion $F$ of every site through the on-site term $\Phi\hf^\dagger_iF_i$ (solid lines). Because the orbitals are not orthogonal, $\{\hf_i,\hf^\dagger_j\}=S_{ij}$, the commutator $[\hf_i,H_{\rm MF}]$ also couples $\hf_i$ to the $F$ of the other sites with the amplitude $S_{ij}\Phi$ [Eq.~\eqref{eq:Xi0}] (dashed lines). Therefore the orthogonal fermions of the other sites, $\Phi\sum_{j\neq i}S_{ij}F_j$, form the bath of the impurity at site $i$. The overlap $S_{ij}$ is denoted $O_{ij}$ in the figure.}
 \label{fig:cF}
\end{figure}

\emph{Mean-field theory of $c$ and $F$.}---The closed pair of Eq.~\eqref{eq:eomat} is equivalent to the mean-field level of the ancilla theory~\cite{ZhangSachdev2020,ZhaoMixed2025,ZhaoAncilla2025,Zhang2026} restricted to the $(c,F)$ subspace. The mean-field theory is (Fig.~\ref{fig:cF})
\begin{equation}
 \begin{aligned}
 H_{\rm MF}&=-\mu\sum_{\bk\alpha}c^\dagger_{\bk\alpha}c_{\bk\alpha}+\sum_{i\alpha}\Big[\Sigma_\infty\hf^\dagger_{i\alpha}\hf_{i\alpha}+\epsilon_FF^\dagger_{i\alpha}F_{i\alpha}+\Phi\big(\hf^\dagger_{i\alpha}F_{i\alpha}+F^\dagger_{i\alpha}\hf_{i\alpha}\big)\Big]\\
 &=\sum_{\bk\alpha}\Big[(S_{\bk}\Sigma_\infty-\mu)\,c^\dagger_{\bk\alpha}c_{\bk\alpha}+\epsilon_F\,F^\dagger_{\bk\alpha}F_{\bk\alpha}+\sqrt{S_{\bk}}\,\Phi\big(e^{-i\theta_{\bk}}c^\dagger_{\bk\alpha}F_{\bk\alpha}+e^{i\theta_{\bk}}F^\dagger_{\bk\alpha}c_{\bk\alpha}\big)\Big],
 \end{aligned}
 \label{eq:hmf}
\end{equation}
where $\mu_{\rm imp}=\mu$ at this order and
\begin{equation}
 \Sigma_\infty=U(\Nf-1)\big(q-\tfrac12\big),\qquad
 \epsilon_F=\Sigma_\infty-\mu+U(2q-1)=U(\Nf+1)\big(q-\tfrac12\big)-\mu,
 \label{eq:epsF}
\end{equation}
that is, $\epsilon_F=qE_+-(1-q)E_-$. In the second line $F_{\bk\alpha}=N_M^{-1/2}\sum_ie^{-i\bk\cdot\bm R_i}F_{i\alpha}$, and $\theta_{\bk}$ is the phase of the form factor, $\mu_f(\bk)=|\mu_f(\bk)|e^{i\theta_{\bk}}$, which winds by $2\pi$ around $\bk_0$: the hybridization carries the vortex of the Chern band. Because $F$ does not disperse in $H_{\rm MF}$, the vortex can be moved into it, $F_{\bk\alpha}\to e^{-i\theta_{\bk}}F_{\bk\alpha}$, which leaves $\epsilon_FF^\dagger_{\bk\alpha}F_{\bk\alpha}$ and the spectrum unchanged and makes the coupling $\sqrt{S_{\bk}}\,\Phi(c^\dagger_{\bk\alpha}F_{\bk\alpha}+{\rm h.c.})$; the transformation is singular only at $\bk_0$, where $S_{\bk}=0$ and the coupling vanishes. We use this gauge below, $F_{\bk\alpha}=e^{-i\theta_{\bk}}N_M^{-1/2}\sum_ie^{-i\bk\cdot\bm R_i}F_{i\alpha}$; the vortex then reappears in the phase $e_{\bk}$ of the bath, Eq.~\eqref{eq:Xi0k}. Integrating out $F$ gives the band electron the Hubbard-I self-energy $\Sigma_c(\bk)=S_{\bk}\Sigma_{\rm at}(z)$. 

\emph{The orthogonal fermion.}---With $\chi_{+\alpha}=\hf_\alpha P_{n_f+1}/\sqrt{1-q}$ and $\chi_{-\alpha}=\hf_\alpha P_{n_f}/\sqrt q$, the Hubbard operators that remove the flavor from the atom with $n_f+1$ and with $n_f$ electrons, normalized to $\langle\{\chi_{\pm\alpha},\chi^\dagger_{\pm\alpha}\}\rangle=1$, we have
\begin{equation}
 \hf^G_\alpha=\sqrt{1-q}\,\chi_{+\alpha}+\sqrt q\,\chi_{-\alpha},\qquad
 F^G_\alpha=\sqrt q\,\chi_{+\alpha}-\sqrt{1-q}\,\chi_{-\alpha},
 \label{eq:psi}
\end{equation}
 $F^G_\alpha$  thus should not be viewed as a bound state. It is essentially a different linear combination of the Hubbard operators, the one orthogonal to the electron, $\langle\{F_\alpha,\hf^\dagger_\alpha\}\rangle=0$ and we call it the orthogonal fermion. These operators are meaningful only in the subspace $P_G$ of the atom with $n_f-1$, $n_f$ and $n_f+1$ electrons. 

\subsection{The bath at first order}
\label{app:bath}
\begin{figure}[t]
 \includegraphics[width=0.4\linewidth]{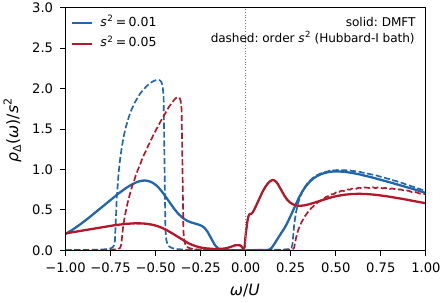}
 \caption{Bath $\rho_\Delta(\omega)/s^2$ at $\nu=-2$ and $T=U/200$ for $s^2=0.01$ and $0.05$, at the same chemical potentials as in Fig.~\ref{fig:mott}(c,d): DMFT (solid) and the bath at order $s^2$, Eq.~\eqref{eq:Deltaexact} with the self-energy of Hubbard-I, evaluated with the distribution $P(S)$ of the lattice (dashed). The closed form, Eq.~\eqref{eq:bath}, is the same bath with $P(S)(1-S)^2=s^2$.}
 \label{fig:bath}
\end{figure}

We simply insert the atomic self-energy into the self-consistency equation, Eq.~\eqref{eq:Dexpand}, and obtain the bath at leading order in $s^2$, $\Delta_f(z)=\zeta\Sigma_{\rm at}(z)\int\dd S\,P(S)(1-S)^2/(\zeta-S\Sigma_{\rm at})$, with $\zeta=z+\mu$. Here the frequency $z$, and $\omega$ on the real axis, is measured from the Fermi level: $\omega=0$ is the Fermi energy. The flat band has no kinetic energy, so that in $H$ of Eq.~\eqref{eq:ham} a band electron without interaction has the energy $\omega=-\mu$, the bare level of the band, and $\zeta=z+\mu$ is the energy measured from this level. For $\nu<0$ the chemical potential is negative and the bare level lies above the Fermi level, at $-\mu\simeq2.3U$ beyond $\nu=-2$. For a Lorentzian orbital weight, $S_{\bk}=w_f(\bk)/W_f$ with $w_f(\bk)=k^2/(k^2+k_*^2)$ and $k=|\bk-\bk_0|$, the weights are distributed as $P(S)=s^2/(1-S)^2$ on $0<S<1$ at this order [Eq.~\eqref{eq:PSlor}], so that $P(S)(1-S)^2=s^2$ and the integral is elementary:
\begin{equation}
 \Delta_f(z)=s^2\zeta\Sigma_{\rm at}\int_0^1\frac{\dd S}{\zeta-S\Sigma_{\rm at}}=-s^2\zeta\,\ln\Big[1-\frac{\Sigma_{\rm at}(z)}{\zeta}\Big],\qquad
 \rho_\Delta(\omega)=s^2\,|\omega+\mu|.
 \label{eq:bath}
\end{equation}
The spectral density comes from the pole of the integrand. On the real axis $\zeta=\omega+\mu+i0^+$ and $\Impart\Sigma_{\rm at}\le0$, so that $-\Impart(\zeta-S\Sigma_{\rm at})^{-1}=\pi\delta(S-S_\omega)/|\Sigma_{\rm at}(\omega)|$, with $S_\omega=(\omega+\mu)/\Sigma_{\rm at}(\omega)$ the weight of the Bloch state whose Hubbard-I energy is $\omega$. For any $P(S)$, then, with the last equality for $P(S)(1-S)^2=s^2$,
\begin{equation}
 \rho_\Delta(\omega)=-\frac1\pi\Impart\Delta_f(\omega)=\frac{(\omega+\mu)\,\Sigma_{\rm at}}{|\Sigma_{\rm at}|}\,P(S_\omega)(1-S_\omega)^2=|\omega+\mu|\,P(S_\omega)(1-S_\omega)^2=s^2 |\omega+\mu|
 \label{eq:rhoPS}
\end{equation}
for $0<S_\omega<1$ and $\rho_\Delta=0$ if $S_\omega$ is outside of the range $(0,1)$. $\omega+\mu$ and $\Sigma_{\rm at}$ have the same sign and $S_\omega>0$. At a nonzero integer filling the Fermi level lies in a gap of the Hubbard-I bands. We have $S_0=\mu/\Sigma_{\rm at}(0)>1$ and so $\rho_\Delta(0)=0$. For example, at the center of the $\nu=-2$ plateau, $\mu=-2U$, $\Sigma_{\rm at}(0)=-U$ and $S_0=2$: the band of $F$ ends at $\epsilon_F=-U/4$ and the upper band starts at $E_+=U/2$. In the doped state $\epsilon_F$ lies above the Fermi level and  the band of $F$ crosses it. For example,  $S_0=0.42$ for $\mu=-2.3U$, and $\rho_\Delta(0)=s^2|\mu|$. Figure~\ref{fig:bath} compares the bath at the $O(s^2)$ order with that of the full DMFT at $\nu=-2$.

\emph{The bath fermions in terms of $F$.}---The DMFT impurity site action may also be written as an Anderson impurity model. We represent the bath by free fermions $b_{l\alpha}$ at the levels $\epsilon_l$, coupled to the orbital $f_i$ with the amplitudes $V_l$, so $H_{\rm imp}=H_{\rm at}+\sum_{l\alpha}[\epsilon_lb^\dagger_{l\alpha}b_{l\alpha}+V_l(\hf^\dagger_\alpha b_{l\alpha}+b^\dagger_{l\alpha}\hf_\alpha)]$. We have the correspondence $\Delta_f(z)=\Delta_\infty+\sum_lV_l^2/(z-\epsilon_l)$, with the static part $\Delta_\infty=\eta_S^2\Sigma_\infty$ of Appendix~\ref{app:dmft}, and
\begin{equation}
 [\hf_\alpha,H_{\rm imp}]=(\Sigma_\infty-\mu_{\rm imp})\hf_\alpha+\Phi F_\alpha+\sum_lV_lb_{l\alpha},
 \label{eq:eomimp}
\end{equation}
so the orbital $f_i$ sees the bath only through the fermion $\sum_lV_lb_{l\alpha}$. In the following we try to give an interpretation of these bath fermions. Consider the original full model, the orbital $f_i$ at $\bm R_0$ obeys the same equation of motion: since $\{\hf_{0\alpha},\hf^\dagger_{j\beta}\}=\delta_{\alpha\beta}S_{0j}$, the charging term of each site $j$ gives, as in the atom [Eq.~\eqref{eq:eomexact}], $[\hf_{0\alpha},\tfrac U2(\hat n_{fj}-\tfrac{\Nf}2)^2]=S_{0j}(\Sigma_\infty\hf_{j\alpha}+\Phi F_{j\alpha})$, with $F_{j\alpha}$ the orthogonal fermion of site $j$, and
\begin{equation}
 [\hf_{0\alpha},H]=(\Sigma_\infty-\mu)\hf_{0\alpha}+\Phi F_{0\alpha}+\Xi_{0\alpha},\qquad
 \Xi_{0\alpha}=\sum_{j\neq0}S_{0j}\big(\Sigma_\infty\hf_{j\alpha}+\Phi F_{j\alpha}\big).
 \label{eq:Xi0}
\end{equation}
Comparing Eqs.~\eqref{eq:eomimp} and \eqref{eq:Xi0}, $\sum_lV_lb_{l\alpha}$ stands for $\Xi_{0\alpha}$, up to one detail. Because the orbitals overlap, $\Xi_{0\alpha}$ is not orthogonal to $\hf_{0\alpha}$ itself: its component along $\hf_{0\alpha}$ is $\langle\{\Xi_{0\alpha},\hf^\dagger_{0\alpha}\}\rangle=\Sigma_\infty\sum_{j\neq0}|S_{0j}|^2=\eta_S^2\Sigma_\infty=\Delta_\infty$ (the $F_j$ give nothing, since $\langle\delta N'_{j\alpha}\rangle=0$). This component acts as an on-site energy of the orbital: it is the static part of the bath, which shifts the level of the impurity from $\mu$ to $\mu_{\rm imp}=\mu-\Delta_\infty$. The rest, $\tilde\Xi_{0\alpha}=\Xi_{0\alpha}-\Delta_\infty\hf_{0\alpha}$, is orthogonal to the orbital and is the dynamical bath $\sum_lV_lb_{l\alpha}$. The bath fermion is thus made of the orthogonal fermions of the other sites, with the amplitudes $S_{0j}\Phi$, and of the $f_j$ orbitals, with $S_{0j}\Sigma_\infty$ (dashed in Fig.~\ref{fig:cF}). 

We now express $\Xi_{0\alpha}$ in momentum space,
\begin{equation}
 \Xi_{0\alpha}=\frac1{\sqrt{N_M}}\sum_{\bk}e_{\bk}\,(1-S_{\bk})\big(\Phi F_{\bk\alpha}+\Sigma_\infty\sqrt{S_{\bk}}\,c_{\bk\alpha}\big),\qquad e_{\bk}=-e^{i\bk\cdot\bm R_0}e^{i\theta_{\bk}},
 \label{eq:Xi0k}
\end{equation}
so only the momenta with $S_{\bk}\neq1$ contribute. Near the Fermi level of the doped state these Bloch states form a small pocket of the lower Hubbard band around $\bk_0$, made almost purely of $F$ (below), so the term of $c_{\bk}$ is small. With the phases $e_{\bk}$ absorbed into $F_{\bk}$, the bath fermions are then the orthogonal fermions of the other sites in the pocket,
\begin{equation}
 \sum_lV_lb_{l\alpha}=\frac{\Phi}{\sqrt{N_M}}\sum_{\bk}\phi_{\bk}F_{\bk\alpha},
 \label{eq:bathF}
\end{equation}
with the form factor $\phi_{\bk}$ given below. 

We focus on the region around $\bk_0$, where $S_{\bk}\ll1$. If we dope into the lower Hubbard band, the pocket is formed almost purely by $F$: to first order in $S_{\bk}$ its energy is $E_{\bk}=\epsilon_F+S_{\bk}\Phi^2/(\epsilon_F+\mu)$, and its weight on the electron is of order $S_{\bk}$. The form factor of Eq.~\eqref{eq:bathF} is the coefficient of $F_{\bk}$ in Eq.~\eqref{eq:Xi0k},
\begin{equation}
 \phi_{\bk}=1-S_{\bk}\simeq\frac{k_*^2}{k^2+k_*^2},
 \label{eq:phik}
\end{equation}
which is $1$ at $\bk_0=0$.  The bath is therefore $\Phi^2$ times the density of states of $F$ counted with $\phi_{\bk}^2$,
\begin{equation}
 \rho_\Delta(\omega)=\Phi^2\rho_F(\omega),\qquad
 \rho_F(\omega)=\frac1{N_M}\sum_{\bk}\phi_{\bk}^2v_{\bk}^2\,\delta(\omega-E_{\bk}),
 \label{eq:rhoF}
\end{equation}
with $v_{\bk}^2$ the weight of $F$ in the band state at $E_{\bk}$, which tends to one around $\bk_0=0$. At the top of the pocket, $\phi\to1$, the band has the density $P(S)\,|\dd S/\dd E|=s^2|\epsilon_F+\mu|/\Phi^2$ with $P(S)\simeq s^2$, and $\rho_\Delta(\epsilon_F)=s^2|\epsilon_F+\mu|$, matching Eq.~\eqref{eq:bath}. 

\subsection{Hubbard-III at first order}
\label{app:h3}
With the bath $\Delta_f(z)$ of Appendix~\ref{app:bath}, the impurity site is not just an atomic Mott insulator, and we must obtain the correction to $\Sig(z)$. In this subsection we show that the self-energy keeps the pole of $F$, now with a self-energy $\Sigma_F$ of its own,
\begin{equation}
 \Sig(z)=\Sigma_\infty+\frac{\Phi^2}{z-\epsilon_F-\Sigma_F(z)},\qquad
 \Sigma_F(z)=(\Nf+1)\big[\Delta_f(z)-\Delta_\infty\big]+O(s^4),
 \label{eq:h3}
\end{equation}
with $\epsilon_F$ and $\Sigma_\infty$ of Eq.~\eqref{eq:epsF}. The second expression is equivalent to Hubbard's third approximation~\cite{Hubbard1964}. In brief, the fermion $F$ at the impurity site decays into a bath fermion combined with a flip of the moment of the impurity atom, through a vertex $U$ of the Kondo form, which gives the scattering rate
\begin{equation}
 \Gamma_F(\omega)\equiv-\Impart\Sigma_F(\omega+i0^+)=(\Nf+1)\,\pi\rho_\Delta(\omega)=(\Nf+1)\,\pi s^2|\omega+\mu| .
 \label{eq:gammapsi}
\end{equation}

\emph{Derivation.}---We use the Anderson model of Eq.~\eqref{eq:eomimp}, $H=H_{\rm at}+H_{\rm bath}+H_{\rm hyb}$, and write $\Delta(z)\equiv\Delta_f(z)-\Delta_\infty=\sum_lV_l^2/(z-\epsilon_l)$ for the dynamical part of the bath, the only part into which an electron can decay. The retarded Green function $\langle\!\langle A;B^\dagger\rangle\!\rangle_z=-i\int_0^\infty\dd t\,e^{izt}\langle\{A(t),B^\dagger\}\rangle$ obeys the equation of motion
\begin{equation}
 z\,\langle\!\langle A;B^\dagger\rangle\!\rangle_z=\langle\{A,B^\dagger\}\rangle+\langle\!\langle[A,H];B^\dagger\rangle\!\rangle_z=\langle\{A,B^\dagger\}\rangle+\langle\!\langle A;[H,B^\dagger]\rangle\!\rangle_z,
 \label{eq:eomR}
\end{equation}
where in the second form the commutator acts on $B^\dagger$; the two forms are equal in a stationary state, in which $\langle[Y,H]\rangle=0$ for every operator $Y$. For the electron, $[\hf_\alpha,H]=(\Sigma_\infty-\mu_{\rm imp})\hf_\alpha+\Phi F_\alpha+\sum_lV_lb_{l\alpha}$, and the bath term gives $\Delta(z)\,G_f$ exactly, since $(z-\epsilon_l)\langle\!\langle b_{l\alpha};\hf^\dagger_\alpha\rangle\!\rangle=V_lG_f$. For $F$ the commutator has three parts, $[F_\alpha,H]=[F_\alpha,H_{\rm at}]+[F_\alpha,H_{\rm bath}]+[F_\alpha,H_{\rm hyb}]$: the atom gives $\Phi\hf_\alpha+\epsilon_FF_\alpha$ in $P_G$ [Eq.~\eqref{eq:eomat}], the bath gives zero because $F_\alpha$ contains no bath operator, and the hybridization gives a new operator,
\begin{equation}
 [F_\alpha,H_{\rm hyb}]\equiv UX_\alpha,\qquad
 X_\alpha=\frac1\Phi\Big(\delta N'_\alpha[\hf_\alpha,H_{\rm hyb}]+[N'_\alpha,H_{\rm hyb}]\,\hf_\alpha\Big).
 \label{eq:Xa}
\end{equation}
Hence
\begin{equation}
 \big[z+\mu_{\rm imp}-\Sigma_\infty-\Delta(z)\big]G_f=1+\Phi\langle\!\langle F_\alpha;\hf^\dagger_\alpha\rangle\!\rangle,\qquad
 (z-\epsilon_F)\langle\!\langle F_\alpha;\hf^\dagger_\alpha\rangle\!\rangle=\Phi G_f+U\langle\!\langle X_\alpha;\hf^\dagger_\alpha\rangle\!\rangle .
\end{equation}
The bath fermion reaches $F$ only through $X_\alpha$. Defining $\Sigma_F$ by $U\langle\!\langle X_\alpha;\hf^\dagger_\alpha\rangle\!\rangle=\Sigma_F\langle\!\langle F_\alpha;\hf^\dagger_\alpha\rangle\!\rangle$, the two equations give the first equation of Eq.~\eqref{eq:h3}. With the measured $\Phi^2=U^2\langle\delta N_\alpha'^2\rangle$ and $\Sigma_\infty$ it holds to all orders, as the first step of a Mori continued fraction~\cite{Mori1965}, and $\tilde G_F(z)=[z-\epsilon_F-\Sigma_F(z)]^{-1}=(\Sig-\Sigma_\infty)/\Phi^2$ is the local propagator of $F$, which CT-HYB gives directly; in the band, where $c_{\bk}$ couples to $F_{\bk}$ with the amplitude $\sqrt{S_{\bk}}\,\Phi$, $G_F(\bk,z)=\tilde G_F(z)(\zeta-S_{\bk}\Sigma_\infty)G_c(\bk,z)$, which reduces to $\tilde G_F$ at the zero of the orbital weight. To find $\Sigma_F$ we apply the second form of Eq.~\eqref{eq:eomR} to $\langle\!\langle X_\alpha;B^\dagger\rangle\!\rangle$ with $B=b_{l\alpha}$, $\hf_\alpha$ and $F_\alpha$, using $[H,B^\dagger]=[B,H]^\dagger$ and the commutators above. Since $X_\alpha$ is orthogonal to $\hf_\alpha$ and to the bath, $\langle\{X_\alpha,b^\dagger_{l\alpha}\}\rangle=\langle\{X_\alpha,\hf^\dagger_\alpha\}\rangle=0$, while $\langle\{X_\alpha,F^\dagger_\alpha\}\rangle=O(V^2)$ only shifts $\epsilon_F$. Hence
\begin{equation}
 \begin{gathered}
 (z-\epsilon_l)\langle\!\langle X_\alpha;b^\dagger_{l\alpha}\rangle\!\rangle=V_l\langle\!\langle X_\alpha;\hf^\dagger_\alpha\rangle\!\rangle,\qquad
 (z+\mu_{\rm imp}-\Sigma_\infty)\langle\!\langle X_\alpha;\hf^\dagger_\alpha\rangle\!\rangle=\Phi\langle\!\langle X_\alpha;F^\dagger_\alpha\rangle\!\rangle+\sum_lV_l\langle\!\langle X_\alpha;b^\dagger_{l\alpha}\rangle\!\rangle,\\
 (z-\epsilon_F)\langle\!\langle X_\alpha;F^\dagger_\alpha\rangle\!\rangle=\Phi\langle\!\langle X_\alpha;\hf^\dagger_\alpha\rangle\!\rangle+U\langle\!\langle X_\alpha;X^\dagger_\alpha\rangle\!\rangle,
 \end{gathered}
\end{equation}
the mirror images of the equations above, with $U\langle\!\langle X_\alpha;X^\dagger_\alpha\rangle\!\rangle$ as the source. The first two give $g(z)\langle\!\langle X_\alpha;\hf^\dagger_\alpha\rangle\!\rangle=\Phi\langle\!\langle X_\alpha;F^\dagger_\alpha\rangle\!\rangle$, with $g(z)=z+\mu_{\rm imp}-\Sigma_\infty-\Delta(z)$, and the third then $\langle\!\langle X_\alpha;F^\dagger_\alpha\rangle\!\rangle=\bar G_F\,U\langle\!\langle X_\alpha;X^\dagger_\alpha\rangle\!\rangle$, with $\bar G_F=[z-\epsilon_F-\Phi^2/g(z)]^{-1}$. On the other side, the equations for $G_f$ and $\langle\!\langle F_\alpha;\hf^\dagger_\alpha\rangle\!\rangle$ give $\langle\!\langle F_\alpha;\hf^\dagger_\alpha\rangle\!\rangle=\Phi G_f/(z-\epsilon_F-\Sigma_F)$ with $G_f^{-1}=g-\Phi^2/(z-\epsilon_F-\Sigma_F)$. Inserting both into the definition $\Sigma_F\langle\!\langle F_\alpha;\hf^\dagger_\alpha\rangle\!\rangle=U\langle\!\langle X_\alpha;\hf^\dagger_\alpha\rangle\!\rangle$ gives
\begin{equation}
 U^2\langle\!\langle X_\alpha;X^\dagger_\alpha\rangle\!\rangle=\frac{\Sigma_F}{1-\bar G_F\Sigma_F}=\Sigma_F+\Sigma_F\bar G_F\Sigma_F+\dots,\qquad
 \bar G_F(z)=\Big[z-\epsilon_F-\frac{\Phi^2}{z+\mu_{\rm imp}-\Sigma_\infty-\Delta(z)}\Big]^{-1},
\end{equation}
so $U^2\langle\!\langle X_\alpha;X^\dagger_\alpha\rangle\!\rangle$ is the full scattering amplitude of $F$, and $\Sigma_F$ its part that does not return through $F$. $X_\alpha$ is of first order in $V$, so $\Sigma_F$ is of second order and the returns are of fourth. At second order we may also drop the hybridization from the motion of the $X$ fermion, since its first-order correction changes the number of bath fermions by one and averages to zero. Hence $\Sigma_F=U^2\Pi+O(V^4)$, with the bubble $\Pi(z)=\langle\!\langle X_\alpha;X^\dagger_\alpha\rangle\!\rangle^{(0)}_z$ of the pair moving without hybridization.

\begin{figure}[h]
 \includegraphics[width=0.92\linewidth]{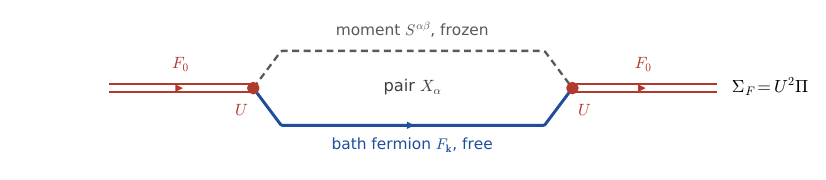}
 \caption{Self-energy of $F$ in Hubbard-III. $F_0$ (red double line) from the impurity site decays with the vertex $U$ (red dots), $[F_\alpha,H_{\rm hyb}]=UX_\alpha$, into the pair $X_\alpha$ of Eq.~\eqref{eq:Xcomp}, a bath fermion (blue) and the local moment (dashed).   }
 \label{fig:h3diag}
\end{figure}

Written out, $X_\alpha$ is a bath fermion times an operator of the atom,
\begin{equation}
 X_\alpha=\frac1\Phi\sum_lV_l\Big[\delta N'_\alpha\,b_{l\alpha}-\sum_{\beta\neq\alpha}\hf^\dagger_\beta\hf_\alpha\,b_{l\beta}-\sum_{\beta\neq\alpha}b^\dagger_{l\beta}\hf_\beta\hf_\alpha\Big].
 \label{eq:Xabc}
\end{equation}
It has three terms and we will drop the third term: it takes the impurity site out of the subspace $P_G$, to $n_f\pm2$ electrons with a cost of about $U$ beyond the Hubbard bands. With the traceless moment $S^{\alpha\beta}=\hf^\dagger_\beta\hf_\alpha-q\delta_{\alpha\beta}$ and $\delta N'_\alpha=-S^{\alpha\alpha}$ in the multiplet, the first two terms  form one composite operator,
\begin{equation}
 X_\alpha=-\frac1\Phi\sum_\beta S^{\alpha\beta}\sum_lV_lb_{l\beta}=-\frac1{\sqrt{N_M}}\sum_{\bk\beta}\phi_{\bk}\,S^{\alpha\beta}F_{\bk\beta},
 \label{eq:Xcomp}
\end{equation}
with the local moment combined with a bath fermion in Eq.~\eqref{eq:bathF} (Fig.~\ref{fig:h3diag}). Without  the hybridization the local moment is frozen and the bath fermion is free, so we have
\begin{equation}
 \Pi(z)=\frac1{\Phi^2}\sum_{l\beta}V_l^2\,\frac{(1-n_l)\langle S^{\alpha\beta}S^{\beta\alpha}\rangle+n_l\langle S^{\beta\alpha}S^{\alpha\beta}\rangle}{z-\epsilon_l},
 \label{eq:sigFb}
\end{equation}
with $n_l$ the occupation of the bath state, and then,
\begin{equation}
 \Pi(z)=\frac{C}{\Phi^2}\,\Delta(z),\qquad C=\sum_\beta\langle S^{\alpha\beta}S^{\beta\alpha}\rangle=\langle(n_\alpha-q)^2\rangle+\sum_{\beta\neq\alpha}\langle n_\beta(1-n_\alpha)\rangle=q(1-q)(\Nf+1).
 \label{eq:bubble}
\end{equation}
In the end we reach $\Sigma_F(z)=U^2\Pi=(\Nf+1)\Delta(z)$.

\begin{figure*}[t]
 \includegraphics[width=0.92\textwidth]{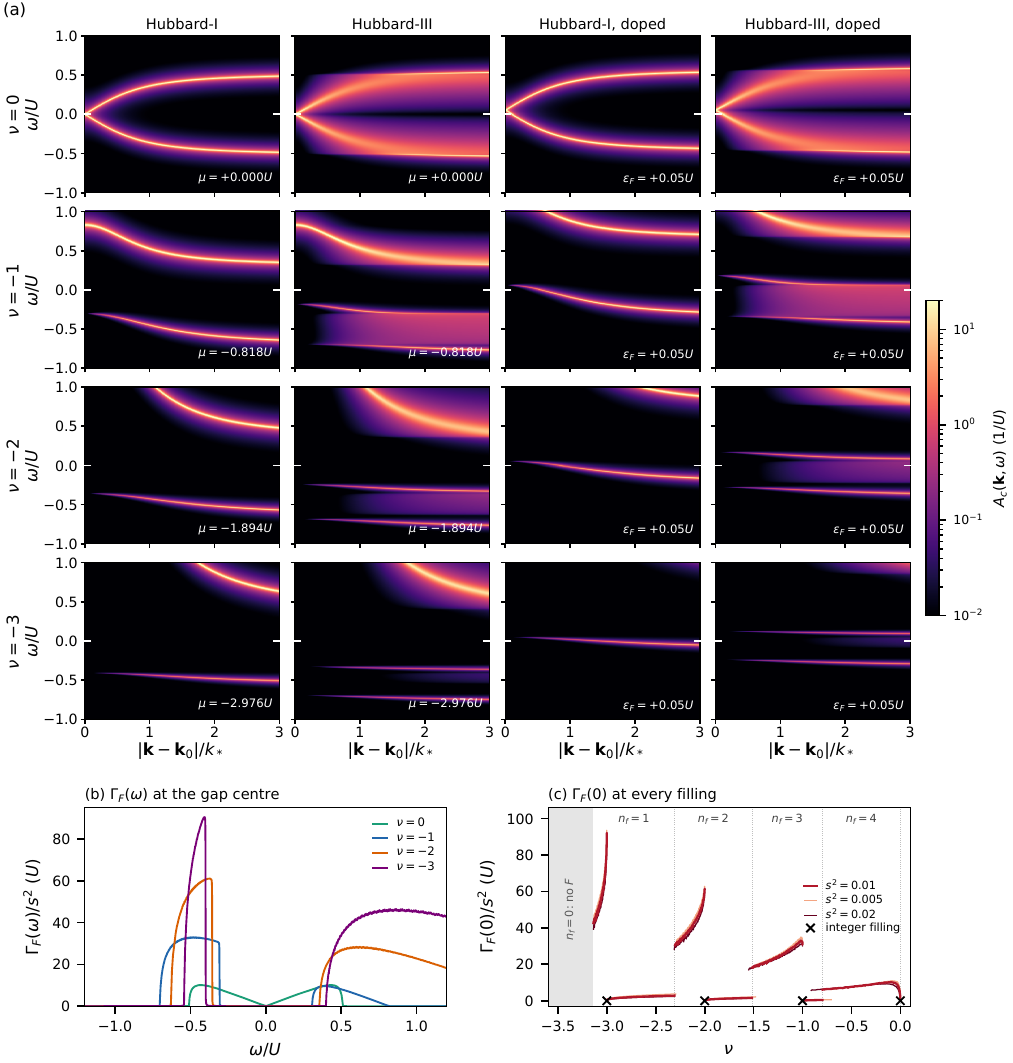}
 \caption{Hubbard-I and Hubbard-III at first order in $s^2$ at the integer fillings ($\Nf=8$, Lorentzian texture, $T=0$). (a) $A_c(\bk,\omega)$ along $|\bk-\bk_0|/k_*$ at $s^2=0.01$ for $\nu=0$, $-1$, $-2$ and $-3$ (rows). The columns show Hubbard-I and Hubbard-III with $\mu$ at the center of the gap, and the state doped away from neutrality with the level of $F$ $0.05U$ above the Fermi level. The color scale is logarithmic, and ticks at the edges mark the Fermi level. The spectra are computed on the real axis with a broadening of $0.004U$, without continuation. The sharp lines bounding the occupied continuum in Hubbard-III are split-off poles. (b) Decay rate of $F$, $\Gamma_F(\omega)/s^2$, at the centers of the gaps. (c) $\Gamma_F(0)/s^2$ at every filling from the exact zone integral, for $s^2=0.01$ (thick) and $0.005$, $0.02$ (thin). The charge $n_f$ of the atom is marked above each pair of branches. Below each integer the Fermi level lies in the band of $F$, and above it the Fermi level lies in the flat bottom of the upper band. Crosses mark $\Gamma_F(0)=0$ at the integers, and the shaded region is $n_f=0$, where there is no $F$.}
 \label{fig:h3int}
\end{figure*}

\emph{Consequences.}---In Fig.~\ref{fig:h3int} we show the spectral function in Hubbard-I and Hubbard-III. At a nonzero integer filling the Fermi level lies in the gap between $\epsilon_F$ and $E_+$, so that $\Gamma_F(0)=0$: Hubbard-III keeps the gap, slightly reduced, and broadens the dispersing parts of the bands in proportion to $|\omega+\mu|$. Between two integers $\Gamma_F(0)$ jumps [Fig.~\ref{fig:h3int}(c)]. There $\Gamma_F(0)=(\Nf+1)\pi\rho_\Delta(0)$ with $\rho_\Delta(0)=|\mu|P(S_0)(1-S_0)^2$, Eq.~\eqref{eq:rhoPS}, where $S_0=\mu/\Sigma_{\rm at}(0)$ is the weight of the Bloch state at the Fermi level; the figure uses the $P(S)$ of the lattice. The simple expression of Eq.~\eqref{eq:gammapsi} is its value for $P(S)(1-S)^2=s^2$, which holds for the dispersing states, $0<S_0<1$, away from the flat region $S\simeq1$. From $\nu=-2$ to $-1$, for example, the atom holds $n_f=2$ electrons up to $\mu=-3U/2$, where its charge states $2$ and $3$ are degenerate, and $n_f=3$ above. With $n_f=2$ the added electrons fill the flat bottom of the upper Hubbard band, $S_0\simeq1$, where $(1-S_0)^2$ and with it the bath vanish, so that $\Gamma_F(0)\simeq0$. With $n_f=3$ the Fermi level lies in the dispersing band of $F$ of the new atom, $0<S_0<1$, and $\Gamma_F(0)$ approaches the simple expression $(\Nf+1)\pi s^2|\mu|$ near $\nu=-1$, where the Fermi level is close to $\bk_0$; it falls below it as the Fermi level moves toward the flat region. At first order the charge of the atom is an integer fixed by $\mu$, so the change is a jump. In the doped state the electron scatters through $F$ at the rate
\begin{equation}
 \Gamma_0\equiv-\Impart\Sig(0)=\frac{\Phi^2\gamma_F}{\epsilon^{*2}_F+\gamma_F^2},
 \label{eq:gamma0}
\end{equation}
with $\gamma_F=\Gamma_F(0)$ and $\epsilon^*_F$ the level of $F$, shifted by the real part of $\Sigma_F$, measured from the Fermi level. It moves with doping. At the integer it lies below the Fermi level, $\epsilon_F=-U/4$ at the center of the $\nu=-2$ plateau. It crosses the Fermi level when the doping starts, at $\mu=-9U/4$. In the doped state the Fermi level lies in the band of $F$ below its top, so that $\epsilon^*_F>0$ grows with the number of holes, up to the depth $\Phi^2/|\mu|\simeq0.08U$ of the pocket ($\epsilon_F=0.05U$ at $\mu=-2.3U$ in Hubbard-I). $F$ is a sharp resonance at the Fermi level when it decays slowly on this scale, $\gamma_F\ll\epsilon^*_F$, and then $\Gamma_0\simeq\Phi^2\gamma_F/\epsilon^{*2}_F$. In the opposite limit, $\gamma_F\gg\epsilon^*_F$, $F$ is overdamped, $\Gamma_0\simeq\Phi^2/\gamma_F$, and the electron scatters less the faster $F$ decays. Since $\epsilon^*_F\to0$ when the doping starts, $F$ is overdamped at small doping for any $s^2$; at $x\simeq0.2$ beyond $\nu=-2$ a sharp $F$ requires $\gamma_F=(\Nf+1)\pi s^2|\mu|\ll\epsilon^*_F\simeq0.05U$, the level of $F$ at $\mu=-2.3U$ in Hubbard-I, that is $s^2\ll10^{-3}$. 

\subsection{Kondo interpretation}
\label{app:kondo}
Eliminating the charge fluctuations to $n_f\pm1$ by a Schrieffer--Wolff transformation~\cite{Schrieffer1966} maps the impurity Anderson model onto a Kondo model, in which the local moment couples to the bath fermions, the orthogonal fermions $F_{\bk}$ of the other sites with the form factor $\phi_{\bk}=1-S_{\bk}$ (Appendix~\ref{app:bath}),
\begin{equation}
 H_K=\frac{J_K}{N_M}\sum_{\bk\bk'}\phi_{\bk}\phi_{\bk'}F^\dagger_{\bk\alpha}\Big(S^{\alpha\beta}+\frac{\epsilon_F}{U}\delta_{\alpha\beta}\Big)F_{\bk'\beta},\qquad J_K=\frac{U\Phi^2}{E_+E_-}.
 \label{eq:kondo}
\end{equation}
This is Eq.~\eqref{eq:jk} of the main text with the potential scattering included. The exchange is of order $U$. But the effective coupling is suppressed by the density of states of the $F$ fermions: $\rho_F=\rho_\Delta/\Phi^2=O(s^2)/U$ [Eq.~\eqref{eq:rhoF}]. The dimensionless coupling
\begin{equation}
 g=\Nf J_K\rho_F(0)=\frac{\Nf U\rho_\Delta(0)}{E_+E_-}
 \label{eq:gdef}
\end{equation}
vanishes at a nonzero integer filling, where $\rho_\Delta(0)=0$, and is $\Nf Us^2|\mu|/(E_+E_-)$ in the doped state. Under the renormalization group $g$ flows as $dg/d\ln D=-g^2$, which gives the Kondo temperature $T_K\simeq De^{-1/g}$, with $D$ the window around the Fermi level in which $\rho_\Delta$ is flat, about the width of the band of $F$.

\emph{Derivation.}---For $V_l\ll E_\pm$ there are two virtual processes: an electron enters the orbital from the bath and leaves again, at the cost $E_+$, or leaves into the bath and comes back, at the cost $E_-$, with $E_\pm$ of Eq.~\eqref{eq:Epm} at the level $\mu_{\rm imp}$. In the multiplet they give
\begin{equation}
 H_K=J_0\sum_{ll'}V_lV_{l'}\,b^\dagger_{l'\alpha}\Big(S^{\alpha\beta}+\frac{\epsilon_F}U\,\delta_{\alpha\beta}\Big)b_{l\beta},\qquad
 J_0=\frac1{E_+}+\frac1{E_-}=\frac U{E_+E_-}.
 \label{eq:sw}
\end{equation}
Both processes add to the exchange, and the potential scattering, $J_0\epsilon_F/U$, vanishes when the level of $F$ is at the Fermi level. With the bath fermions of Eq.~\eqref{eq:bathF} this is Eq.~\eqref{eq:kondo}, with $J_K=J_0\Phi^2$.

\emph{Hubbard-III as the leading order of Kondo scattering.}---The Kondo coupling scatters a bath fermion into the same pair as the decay of $F$, $[F_{\bk\alpha},H_K]=-(J_K\phi_{\bk}/\sqrt{N_M})X_\alpha$ up to potential scattering [Eq.~\eqref{eq:Xcomp}], with the vertex $J_K\phi_{\bk}/\sqrt{N_M}$ in place of $U$. At second order, with the moment frozen, its scattering amplitude contains the same bubble $\Pi$ as Eq.~\eqref{eq:bubble},
\begin{equation}
 T^{(2)}_{\bk\bk'}(z)=\frac{\phi_{\bk}\phi_{\bk'}}{N_M}\,J_K^2\Big[\Pi(z)+\Big(\frac{\epsilon_F}U\Big)^2\frac{\Delta(z)}{\Phi^2}\Big],
 \label{eq:TBorn}
\end{equation}
and when the level of $F$ is at the Fermi level, where $J_K=U$ and $\epsilon_F=0$, the Born rate of a bath fermion equals the Hubbard-III decay rate of $F$, $(\Nf+1)\pi\rho_\Delta(0)$. The factors match as follows. The vertex $J_K\phi_{\bk}/\sqrt{N_M}$ of the bath fermion takes the place of the vertex $U$ of $F$ [Eq.~\eqref{eq:Xa}], and both scatter into the same pair $X_\alpha$, so the bubble $\Pi$ is common. Every site carries a moment, and summing the forward amplitude over the $N_M$ sites gives the self-energy of the bath fermion,
\begin{equation}
 \Sigma_F(\bk,z)=N_MT^{(2)}_{\bk\bk}(z)=\phi_{\bk}^2\Big(\frac{J_K}U\Big)^2\Big[\Sigma_F(z)+\epsilon_F^2\frac{\Delta(z)}{\Phi^2}\Big],
\end{equation}
with $\Sigma_F=U^2\Pi=(\Nf+1)\Delta(z)$ the Hubbard-III self-energy of $F$ at the impurity site [Eq.~\eqref{eq:bubble}]. At $J_K=U$ and $\epsilon_F=0$ it is $\phi_{\bk}^2\Sigma_F(z)$: near $\bk_0$, where $\phi_{\bk}=1-S_{\bk}\simeq1$ and $F_{\bk}$ has the local propagator $\tilde G_F$, the bath fermion has the self-energy of $F$ at the impurity site. The decay of $F$ in Hubbard-III and the Born scattering of the bath fermions by the moments are thus the same process. Because the Kondo effect is not perturbative, Hubbard-III must break down at larger $g$; in particular, the screening of the moments below $T_K$ lies beyond any expansion in $s^2$. We therefore rely on the full DMFT solution, which departs qualitatively from Hubbard-III once $s^2$ exceeds about $0.02$.

\section{DMFT results: cascade and Mott transition}
\label{app:results}

\subsection{Cascade}

\begin{figure*}[t]
 \includegraphics[width=\textwidth]{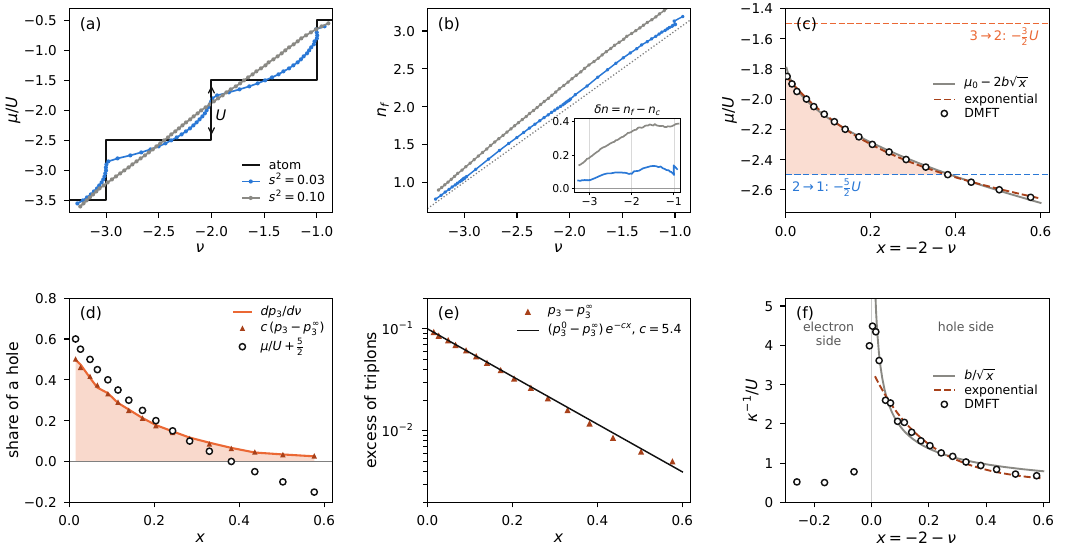}
 \caption{The cascade beyond $\nu=-2$ at $T=U/50$. (a) Chemical potential against filling: the atomic staircase and DMFT at $s^2=0.03$ (Mott-like at this temperature) and $0.10$ (metal). (b) Charge of the orbital, $n_f$, against filling. (c) $\mu$ against the hole doping $x$ at $s^2=0.03$ (circles), with the fit $\mu_0-2b\sqrt x$ (gray) and Eq.~\eqref{eq:muexp} (red dashed). (d) Share of each hole that removes a triplon, $\dd p_3/\dd\nu$, with $\mu/U+\tfrac52$ [Eq.~\eqref{eq:mup3}] and $c\,(p_3-p_3^\infty)$ [Eq.~\eqref{eq:p3law}]. (e) Excess of triplons, $p_3-p_3^\infty$, and $(p_3^0-p_3^\infty)e^{-cx}$. (f) $\kappa^{-1}$ on both sides of $\nu=-2$, with $b/\sqrt x$ and Eq.~\eqref{eq:muexp}.}
 \label{fig:cascademech}
\end{figure*}

For small $s^2$ the inverse compressibility shows a cascade (Fig.~\ref{fig:cascade}). Around each nonzero integer filling it is asymmetric between the electron- and hole-doped sides. At $\nu=-2+x$, $\kappa^{-1}$ drops sharply from its peak at $x=0$. At $\nu=-2-x$ it instead decreases gradually, as $\kappa^{-1}\simeq b/\sqrt x$. Correspondingly, $\mu$ follows $\mu_0-2b\sqrt x$ beyond $\nu=-1$, $-2$ and $-3$ at $s^2=0.03$ and $0.05$ [Fig.~\ref{fig:cascademech}(c) for $\nu=-2$]. In MATBG the same behavior was attributed to a Dirac revival~\cite{Zondiner2020}, whose linear dispersion gives a density of states linear in energy, hence $\delta \mu\propto-\sqrt x$ and $\kappa^{-1}\propto1/\sqrt x$. Our model, however, has no rigid band. In the bad-metal regime the doped holes enter an incoherent continuum without a Fermi surface, so there is no Dirac fermion.

\emph{Mixed-valence Mott insulator.}---The asymmetry and the slow decrease of $\kappa^{-1}$ with hole doping follow instead from the picture of the mixed-valence Mott insulator~\cite{ZhaoMixed2025,ZhaoAncilla2025}. At $s^2=0$ every orbital holds an integer charge $n$, with the interaction energy $E_n=\tfrac U2(n-\tfrac\Nf2)^2$. A hole that takes an orbital from $n$ to $n-1$ electrons costs $\mu=E_n-E_{n-1}$. Between two consecutive integer fillings every hole does the same, so $\mu(\nu)$ is flat there and $\kappa^{-1}$ vanishes. At each integer $\nu$, where $n_f$ jumps by $1$, $\mu$ jumps by $U$. With $s^2>0$ the state at $\nu=-2$ is a mixed-valence Mott state. Besides the doublons, $n_f=2$, it holds triplons, $n_f=3$, and fewer singlons, $n_f=1$. In the following we take $s^2=0.03$. We define $\delta n\equiv n_f-n_c$, the excess of the charge of the impurity site over the band filling. At this $s^2$ it is nearly constant [inset of Fig.~\ref{fig:cascademech}(b)].

Beyond $\nu=-2$ a hole can either turn a triplon into a doublon, $3\to2$, at the price $E_3-E_2=-\tfrac32U$, or a doublon into a singlon, $2\to1$, at $E_2-E_1=-\tfrac52U$. Writing the interaction energy as $\sum_np_nE_n$ over the valences $1$, $2$ and $3$, with $n_f=n_c+\delta n$, we obtain
\begin{equation}
\mu=-\frac52U+U\frac{\dd p_3}{\dd\nu}-\frac52U\frac{\dd\,\delta n}{\dd\nu},
\label{eq:mup3}
\end{equation}
where the last term is small at the doping considered. The price of a hole thus exceeds $-\tfrac52U$ by $U$ times the share of holes that remove a triplon, $\dd p_3/\dd\nu$. Triplons are cheaper to remove, so the first holes deplete them, at a rate set by the density of the triplons left [Fig.~\ref{fig:cascademech}(d)],
\begin{equation}
\frac{\dd p_3}{\dd\nu}=c\,(p_3-p_3^\infty),
\label{eq:p3law}
\end{equation}
where $c$ and $p_3^\infty$, the density of triplons that the bath sustains away from the integer, are fitted to the DMFT results. The law holds over the whole range of doping considered, and at $s^2=0.05$, with nearly the same $c$, over a smaller range. The excess of triplons therefore decays exponentially, $p_3-p_3^\infty=(p_3^0-p_3^\infty)e^{-cx}$, where $p_3^0$ is the density of triplons at $\nu=-2$ [Fig.~\ref{fig:cascademech}(e)]. Inserting this decay into Eq.~\eqref{eq:mup3}, and describing the last term, together with the valences $0$ and $4$ left out there, by a constant background $\kappa_0^{-1}$, we obtain
\begin{equation}
\begin{aligned}
\mu&=\mu_0-\kappa_0^{-1}x-cU(p_3^0-p_3^\infty)\big(1-e^{-cx}\big),\\
\kappa^{-1}&=\kappa_0^{-1}+c^2U(p_3^0-p_3^\infty)\,e^{-cx},
\end{aligned}
\label{eq:muexp}
\end{equation}
where the amplitude $cU(p_3^0-p_3^\infty)$ is fixed by the triplon density, so that only $\mu_0$ and $\kappa_0^{-1}$ are fitted to $\mu$. The fitted background $\kappa_0^{-1}$ is close to $\kappa^{-1}$ on the electron side of $\nu=-2$. Equation~\eqref{eq:muexp} reproduces $\mu$ closely [Fig.~\ref{fig:cascademech}(c)]. Thus $\mu$ falls fast while the triplons are removed and then slowly. The area between $\mu$ and $-\tfrac52U$ is $U$ times the density of triplons removed. The jump of $\mu$ at the integer is thereby moved to its hole side, which gives the asymmetric teeth of $\kappa^{-1}$ [Fig.~\ref{fig:cascademech}(f)]. In this range of doping the exponential form is nearly indistinguishable from the square-root form [Fig.~\ref{fig:cascademech}(c,f)]. The mixed-valence picture therefore explains the observed $\kappa^{-1}$ without a Dirac revival.

\begin{figure*}[t]
 \includegraphics[width=\textwidth]{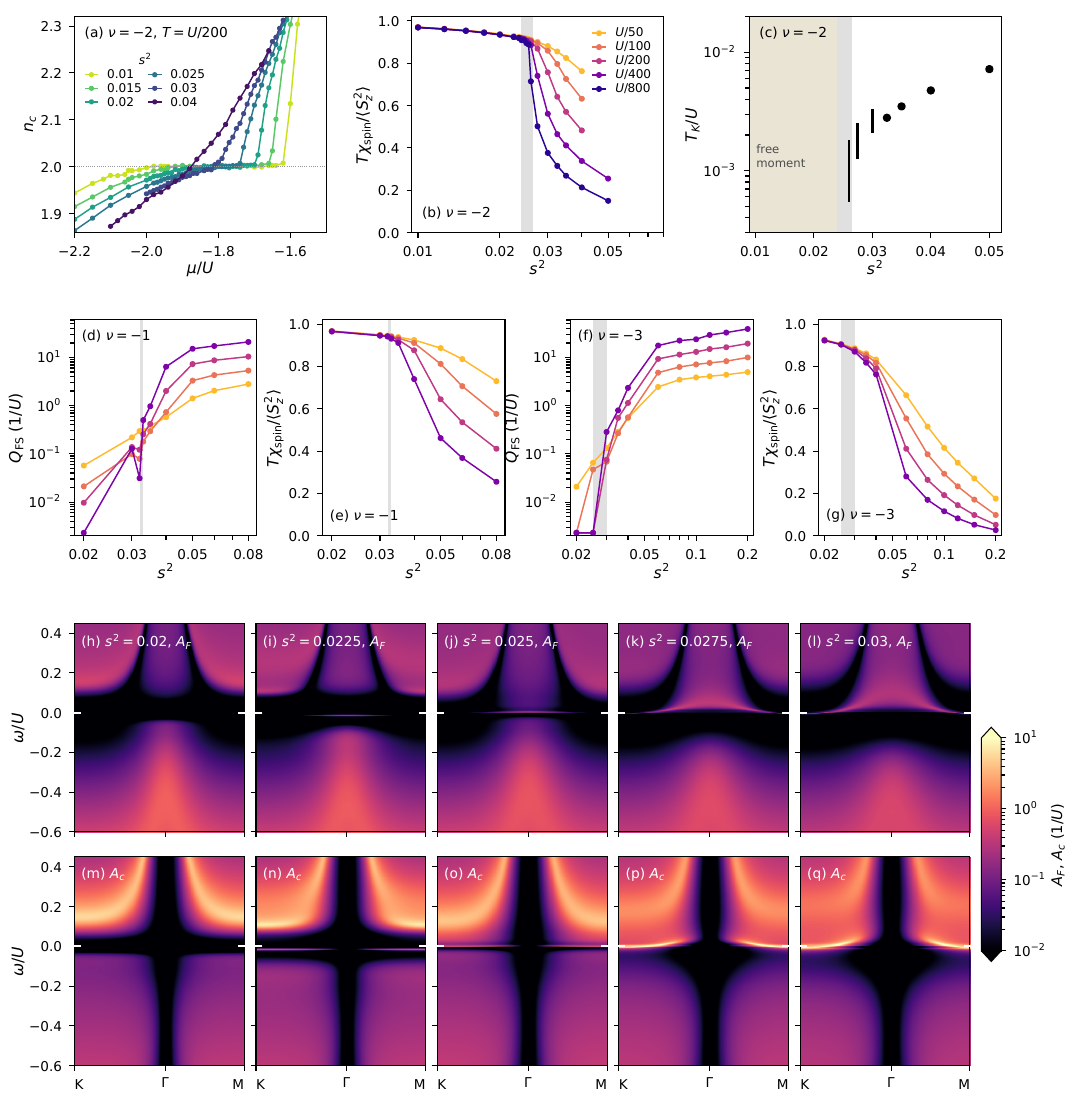}
 \caption{Mott transitions at integer filling. (a) Band filling against $\mu$ at $\nu=-2$ and $T=U/200$. The width of the plateau $n_c=2$ is the charge gap of Fig.~\ref{fig:mott}(a). (b) Unscreened fraction of the moment, $T\chi_{\rm spin}/\langle S_z^2\rangle$, at the center of the plateau, $\mu=-1.80U$, against $s^2$ for the temperatures in the legend. (c) Kondo temperature at $\nu=-2$. Circles mark the points where $\tau=\langle S_z^2\rangle/\chi_{\rm spin}-T$ has saturated (Appendix~\ref{app:tk}). Near the transition (bars), $\tau$ still rises at the lowest temperature. In this case $\tau$ grows on cooling toward $T_K$ while $\tau+T=\langle S_z^2\rangle/\chi_{\rm spin}$ falls toward it, so $T_K$ lies between the two values at that temperature. (d--g) Weight within about $2T$ of the Fermi level on the Fermi surface, $Q_{\rm FS}=\max_{\bk}[-\beta G_c(\bk,\beta/2)/\pi]$, and $T\chi_{\rm spin}/\langle S_z^2\rangle$ against $s^2$ at $\nu=-1$ and $-3$, for the temperatures of (b). (h--q) $A_F(\bk,\omega)$ [Eq.~\eqref{eq:gFk}] and $A_c(\bk,\omega)$ along $K$--$\Gamma$--$M$ at $\nu=-2$ and $T=U/800$ for five $s^2$ across the transition, from the self-energy continued by maximum entropy as in Fig.~\ref{fig:mott}(c,d).}
 \label{fig:mottall}
\end{figure*}

\subsection{Mott transitions at integer filling}
\label{app:mott}

At a non-zero integer filling there is a Mott transition tuned by $s^2$. We illustrate it at $\nu=-2$ with only the Hubbard $U$. At fixed temperature $n_c(\mu)$ has a plateau at $n_c=2$ [Fig.~\ref{fig:mottall}(a)]. Its width gives the charge gap of Fig.~\ref{fig:mott}(a). On the insulating side the spin moment at the center of the plateau is free. Its unscreened fraction $T\chi_{\rm spin}/\langle S_z^2\rangle$ stays close to one and does not depend on temperature [Fig.~\ref{fig:mottall}(b)]. On the metallic side the moment is screened, with a Kondo temperature (Appendix~\ref{app:tk}) that rises continuously from zero near the transition [Fig.~\ref{fig:mottall}(c)]. Sweeps in $s^2$, upward from the insulator and downward from the metal, give the same solution at the lowest temperatures, down to $T=U/800$. A first-order transition, if any, would therefore lie below this temperature. We place the transition at $s^2_c\simeq0.025$. Similar transitions occur at $\nu=-1$ and $-3$, at $s^2_c\simeq0.033$ and $0.027$. There we locate them where the curves of the weight on the Fermi surface at two low temperatures cross [Fig.~\ref{fig:mottall}(d--g)].

\emph{The level of $F$ reaches the Fermi level.}---We obtain the spectral function of the $F$ fermion from the DMFT self-energy alone. In the ancilla picture of Eq.~\eqref{eq:dysonF} the electron $c_{\bk}$ couples to $F_{\bk}$ with the amplitude $\sqrt{S_{\bk}}\,\Phi$. The self-energy is then $\Sig=\Sigma_\infty+\Phi^2\tilde G_F$, where $\tilde G_F=1/(z-\epsilon_F-\Sigma_F)$ is the local propagator of $F$. Eliminating $c_{\bk}$ gives the lattice propagator of $F$,
\begin{equation}
 G_F(\bk,z)=\frac{1}{z-\epsilon_F-\Sigma_F(z)-\dfrac{S_{\bk}\Phi^2}{z+\mu-S_{\bk}\Sigma_\infty}}=\tilde G_F(z)\,\frac{z+\mu-S_{\bk}\Sigma_\infty}{z+\mu-S_{\bk}\Sig(z)},
 \label{eq:gFk}
\end{equation}
with $\tilde G_F=(\Sig-\Sigma_\infty)/\Phi^2$. The spectral function $A_F(\bk,\omega)=-\Impart G_F(\bk,\omega)/\pi$ therefore needs only $\Sig(\omega)$ continued to the real axis, as $A_c$ does. We continue $\Sig$ by maximum entropy~\cite{Jarrell1996}, so the spectra in this paper are qualitative. At $\Gamma$, where $S_{\bk}=0$, the electron decouples and $G_F=\tilde G_F$. There the top of the occupied band is dominated by $F$. The spectra at low temperature show how the gap closes [Fig.~\ref{fig:mottall}(h--q)]. The top of this band lies below the Fermi level in the insulator, touches it at $s^2_c$ and crosses it beyond. In the metal the moments are screened, and the electron acquires a narrow band at the Fermi level away from $\Gamma$. This band forms the heavy Fermi liquid.

\section{DMFT results: the doped Mott state}
\label{app:doped}

\begin{figure*}[t]
 \includegraphics[width=\textwidth]{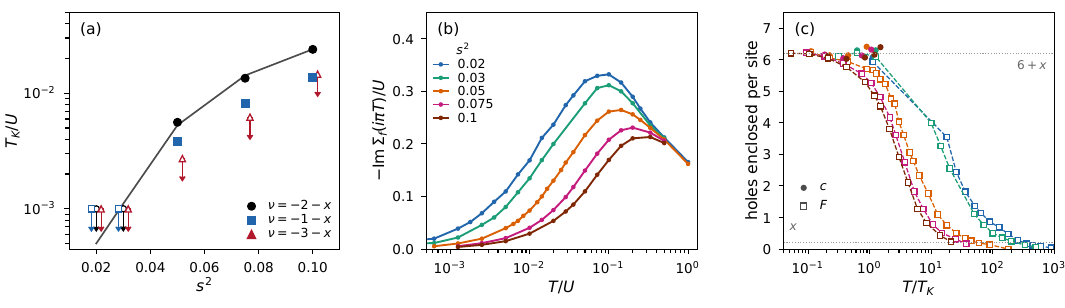}
 \caption{The doped Mott state ($\Nf=8$, $x\simeq0.2$ holes beyond $\nu=-2$, from $T=U$ to $U/3200$). (a) Kondo temperature versus $s^2$ beyond $\nu=-2$, $-1$ and $-3$. For the open symbols with arrows $\tau$ has not saturated at the lowest temperature, and $T_K$ lies below the value shown. Values below $U/1000$ are drawn at $U/1000$. The line is $T_K=De^{-1/g}$ for $\nu=-2-x$ with $D=0.079U$. (b) Electron scattering rate $-\Impart\Sig(i\pi T)$. (c) Holes per site enclosed by the maximum of the weight of the electron (filled) and of $F$ (open) at the Fermi level. Above $T_K$ the maximum of $F$ lies on a band above the Fermi level and is not a Fermi surface. Colors of $s^2$ as in (b).}
 \label{fig:doped}
\end{figure*}

We dope $\nu=-2$ by $x\simeq0.2$ holes and follow the state from $T=U$ down to $U/3200$ for several $s^2$ (Fig.~\ref{fig:doped}). We first describe the DMFT results directly and then fit them with the ancilla theory.

\subsection{Kondo temperature}
\label{app:tk}

We define $T_K=\langle S_z^2\rangle/\chi_{\rm spin}(T\to0)$, with $\langle S_z^2\rangle$ and $\chi_{\rm spin}$ the flavor-traceless parts of the equal-time correlator $\langle n_\alpha n_\beta\rangle$ and of the static susceptibility. We obtain it from $\tau(T)=\langle S_z^2\rangle/\chi_{\rm spin}-T$. If $\chi_{\rm spin}=\langle S_z^2\rangle/(T+T_K)$, of Curie--Weiss form, then $\tau=T_K$ at every temperature. In the doped metal $\tau$ decreases on cooling and levels off at $T_K$, where $\chi_{\rm spin}$ saturates in the Fermi liquid. If $\tau$ has not saturated at the lowest temperature, it is an upper limit of $T_K$. Near the Mott transition at $\nu=-2$, where the moment is almost free at high temperature, $\tau$ instead rises on cooling, and we bracket $T_K$ [Fig.~\ref{fig:mottall}(c)].

Beyond $\nu=-2$ the Kondo temperature follows the one-loop law $T_K=De^{-1/g}$ of Appendix~\ref{app:kondo}, with $g=\Nf U\rho_\Delta(0)/(E_+E_-)$ and the bath $\rho_{\Delta}(0)$ taken at $T_K$. The fitted cutoff $D$ is close to the width $\Phi^2/|\mu|$ of the band of $F$ [Fig.~\ref{fig:doped}(a)]. Beyond $\nu=-1$ and $-3$ the Kondo temperature is lower. 

\subsection{Scattering rate and Fermi surface}
\label{app:dmftE}

\emph{Scattering rate.}---The electron scattering rate $-\Impart\Sig(i\pi T)$ grows on cooling from $T=U$, peaks and then falls [Fig.~\ref{fig:doped}(b)]. Hubbard-III, at first order in $s^2$ with the same impurity charge, reproduces it above about $U/5$ and fails below. Below its maximum the electron rate falls linearly in temperature. At $s^2=0.05$, $\Gamma_0=-\Impart\Sig(i0^+)=(3.4$--$3.8)\,T$ at temperatures between three and nine times $T_K=U/178$.

\emph{Fermi-surface area.}---The weight at the Fermi level, $Q_\beta(S)=-\beta G(S,\beta/2)/\pi$, needs no analytic continuation and can be used as a measure of the ``Fermi surface''. We evaluate it for the electron, $G_c=1/(z+\mu-S\Sig)$, and for $F$, with the propagator of Eq.~\eqref{eq:gFk} at orbital weight $S$. A maximum at $S_{\rm ring}$ indicates a possible Fermi surface, the contour $S_{\bk_F}=S_{\rm ring}$. Its volume is $A_{\rm FS}=\Nf\int_0^{S_{\rm ring}}P(S)\,\dd S$ holes per site. We expect $A_{\rm FS}$ to be $x$ for a small Fermi surface and $\Nf-n_c$ for the conventional Fermi liquid. Indeed we find $A_{\rm FS}=6+x$ below $T_K$, as expected for the HFL [Fig.~\ref{fig:doped}(c)]. Above $T_K$ there is no Fermi surface. The electron weight at the Fermi level has no maximum inside the zone. The maximum of the weight of $F$ lies on its band above the Fermi level, and this band comes down on cooling. At no temperature is there a sharp small Fermi surface. If we nevertheless take the maximum of the weight of $F$ as a Fermi surface, its area grows continuously on cooling [Fig.~\ref{fig:doped}(c)], a further sign that the bad metal has no Fermi surface.

\subsection{Fit with the ancilla theory}
\label{app:spinon}

Here we show that the ancilla framework, in which $c_{\bk}$ hybridizes with a pair of auxiliary fermions $F$ and $F'$, captures the evolution from the bad metal above $T_K$ to the HFL below $T_K$.

\begin{figure*}[t]
 \includegraphics[width=\textwidth]{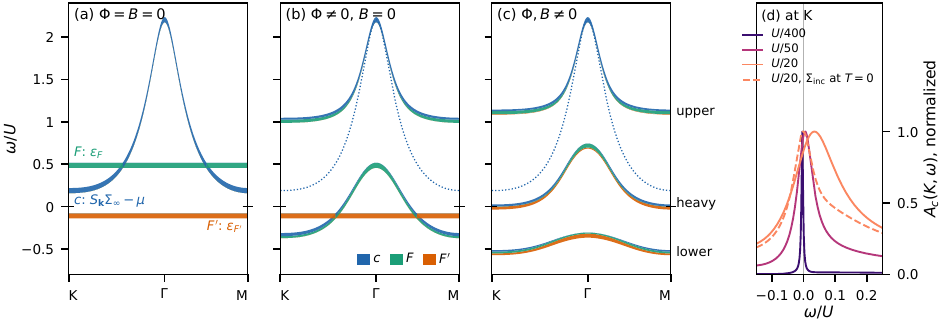}
 \caption{Three bands of the ancilla theory ($s^2=0.05$, $x\simeq0.2$ beyond $\nu=-2$, parameters of Eq.~\eqref{eq:chain} fitted at $U/400$). (a--c) Bands of Eq.~\eqref{eq:hchain} without the couplings, with $\Phi$ only, and with $\Phi$ and $B$. Each band is drawn as a ribbon of fixed width, divided into the weights of $c$ (blue), $F$ (green) and $F'$ (orange). The dotted line is the Hartree energy $S_{\bk}\Sigma_\infty-\mu$ of the electron. (d) $A_c(K,\omega)$ of the ancilla theory at $U/400$, $U/50$ and $U/20$, normalized to its maximum. All parameters are frozen at their values fitted at $U/400$, except three that follow the temperature: the spinon width $\Gamma_{F'}=2.3T$, the chemical potential $\mu$, set by the filling, and the cutoff $a=(\Omega_c^2+\pi^2T^2)^{1/2}$ of $\Sigma_{\rm inc}$ in Eq.~\eqref{eq:siginc}, whose $\gamma_\infty$ and $\Omega_c$ stay frozen. The dashed line is $U/20$ with $a=\Omega_c$, that is, with $\Sigma_{\rm inc}$ at its $T=0$ form. }
 \label{fig:chainbands}
\end{figure*}

\emph{Mean-field theory.}---Without scattering the theory is quadratic. For each flavor
\begin{equation}
 H_{\rm MF}=\sum_{\bk}\Big[(S_{\bk}\Sigma_\infty-\mu)\,c^\dagger_{\bk}c_{\bk}+\sqrt{S_{\bk}}\,\Phi\,\big(c^\dagger_{\bk}F_{\bk}+{\rm H.c.}\big)+\epsilon_FF^\dagger_{\bk}F_{\bk}+\epsilon_{F'}F'^\dagger_{\bk}F'_{\bk}+B\,\big(F^\dagger_{\bk}F'_{\bk}+{\rm H.c.}\big)\Big],
 \label{eq:hmfchain}
\end{equation}
where $F_{\bk}$ and $F'_{\bk}$ are the Fourier components of the local ancillas. In the basis $(c_{\bk},F_{\bk},F'_{\bk})$ the Bloch Hamiltonian is
\begin{equation}
 H_{\bk}=\begin{pmatrix} S_{\bk}\Sigma_\infty-\mu & \sqrt{S_{\bk}}\,\Phi & 0\\ \sqrt{S_{\bk}}\,\Phi & \epsilon_F & B\\ 0 & B & \epsilon_{F'}\end{pmatrix}
 \label{eq:hchain}
\end{equation}
with energies from the Fermi level. Figure~\ref{fig:chainbands} builds its bands with the parameters fitted at $U/400$ for $s^2=0.05$, which place $F$ above and $F'$ below the Fermi level. Without the couplings [Fig.~\ref{fig:chainbands}(a)] the electron disperses at its Hartree energy $S_{\bk}\Sigma_\infty-\mu$, and $F$ and $F'$ are flat levels. The coupling $\Phi$ hybridizes $c$ with $F$ into the two bands of Hubbard-I, while the spinon $F'$ stays at a flat level, representing an unscreened moment [Fig.~\ref{fig:chainbands}(b)]. The coupling $B$ hybridizes $F$ with $F'$ and gives three bands [Fig.~\ref{fig:chainbands}(c)]. The lower and upper bands carry almost all of the weight of $F$ at $K$. They are the Hubbard bands, which $\Sigma_{\rm inc}$ broadens in DMFT. The middle band is the heavy band. Near $K$ it is the combination $Bc-\sqrt{S_{\bk}}\,\Phi F'$. Toward $\Gamma$ it loses the electron and becomes the antibonding level of $F$ and $F'$, so that it is flat and heavy only near the Brillouin zone boundary, where it crosses the Fermi level and forms the large Fermi surface.

\begin{figure*}[t]
 \includegraphics[width=\textwidth]{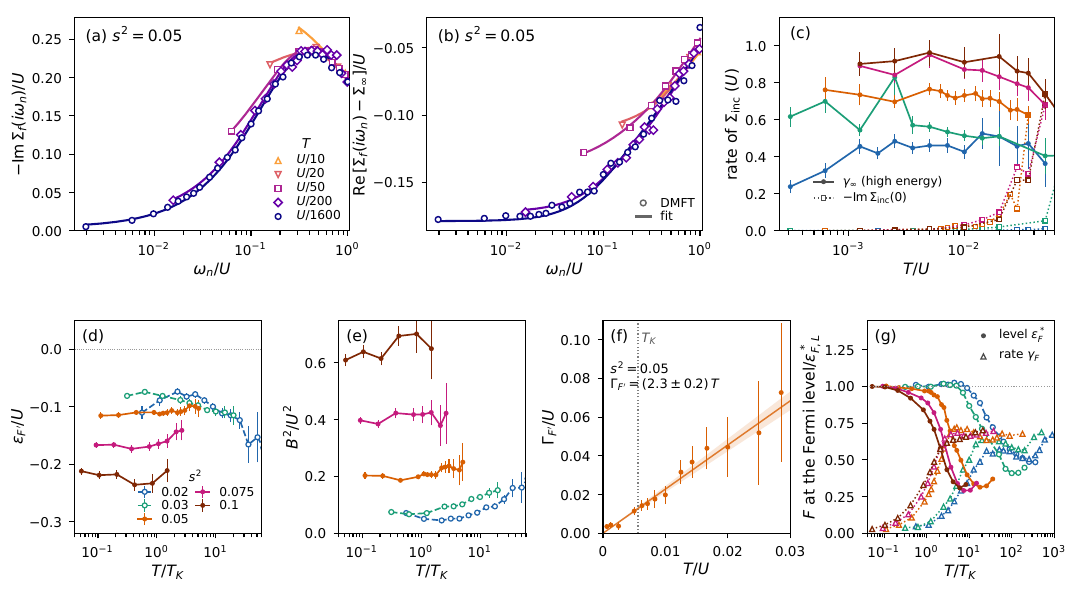}
 \caption{Quality of the fit of Eq.~\eqref{eq:chain} ($x\simeq0.2$ beyond $\nu=-2$, $\Nf=8$). (a) $-\Impart\Sig(i\omega_n)$ and (b) $\Repart[\Sig(i\omega_n)-\Sigma_\infty]$ at $s^2=0.05$ for the temperatures in the legend. Symbols are DMFT, and lines are the ancilla fit. (c) Decay rate of $\Sigma_{\rm inc}$ at high energy, $\gamma_\infty$ (filled circles), and at the Fermi level, $-\Impart\Sigma_{\rm inc}(0)=\gamma_\infty\pi^2T^2/(\Omega_c^2+\pi^2T^2)$ (open squares), for the $s^2$ of the legend in (d). (d) Spinon level $\epsilon_{F'}$ and (e) its residue $B^2$ against $T/T_K$. For $s^2=0.02$ and $0.03$ (open symbols, dashed lines) $T_K$ is an upper limit and $T/T_K$ a lower limit. (f) Scattering rate $\Gamma_{F'}$ of the spinon $F'$ against temperature at $s^2=0.05$. The dotted vertical line indicates $T_K$. Bars in (c--f) are one standard deviation from the Jacobian of the fit. (g) Level $\epsilon^*_F$ (circles) and decay rate $\gamma_F$ (triangles) of $F$ at the first Matsubara frequency, in units of the Luttinger level $\epsilon^*_{F,L}$.}
 \label{fig:chainfit}
\end{figure*}

\emph{Self-energy as a continued fraction.}---Integrating out $F'$ and then $F$ gives the self-energy of the electron as a continued fraction. Scattering enters as the decay rate $\Gamma_{F'}$ of the spinon and the self-energy $\Sigma_{\rm inc}$ of $F$, so that Eqs.~\eqref{eq:dysonF} and \eqref{eq:chain} give
\begin{equation}
 \Sig(z)=\Sigma_\infty+\cfrac{\Phi^2}{z-\epsilon_F-\Sigma_{\rm inc}(z)-\cfrac{B^2}{z-\epsilon_{F'}+i\Gamma_{F'}}}.
 \label{eq:sigcf}
\end{equation}
Without $\Gamma_{F'}$ and $\Sigma_{\rm inc}$ the poles of $G_c(\bk,z)=1/(z+\mu-S_{\bk}\Sig(z))$ are the three bands of Eq.~\eqref{eq:hchain}. The damping broadens them. We fit Eq.~\eqref{eq:sigcf} to the DMFT self-energy at each temperature, with $\Sigma_\infty$ and $\Phi^2$ measured [Eq.~\eqref{eq:eomexact}] and six parameters fitted: the levels $\epsilon_F$ and $\epsilon_{F'}$, the coupling $B^2$, the spinon width $\Gamma_{F'}$, and the rate $\gamma_\infty$ and cutoff $\Omega_c$ of $\Sigma_{\rm inc}$ [Eq.~\eqref{eq:siginc}]. The fit agrees well with DMFT. From $U$ to $U/3200$ it reproduces both parts of the self-energy with an rms deviation of $0.7$--$8\%$ [Fig.~\ref{fig:chainfit}(a,b)], and its parameters change smoothly with temperature [Fig.~\ref{fig:chainfit}(c--f)]. The spinon $F'$ level lies below the Fermi level, between about $-0.07U$ and $-0.25U$ for $\epsilon_{F'}$, and its width is linear in temperature, $\Gamma_{F'}=(2.3\pm0.2)T$ at $s^2=0.05$, which gives the electron the rate $(\Phi^2/B^2)\Gamma_{F'}\simeq3.7T$. At the lowest temperature the level of $F$ at the Fermi level, $\epsilon^*_F=\epsilon_F+\Repart\Sigma_F(0)$, equals the value $\epsilon^*_{F,L}$ that a Fermi surface of the Luttinger volume requires, and its decay rate vanishes [Fig.~\ref{fig:chainfit}(g)].

\emph{Incoherent part.}---The damping $\Sigma_{\rm inc}$ of Eq.~\eqref{eq:chain} is not derived. We design it to have a constant decay rate at high energy, as in Hubbard-III,  and a decay rate proportional to $\omega^2+\pi^2T^2$ at low energy, so that the quasiparticles are undamped at $T=0$. The simplest such function is a constant with a Lorentzian dip,
\begin{equation}
 -\Impart\Sigma_{\rm inc}(\omega)=\gamma_\infty\Big[1-\frac{\Omega_c^2}{\omega^2+a^2}\Big],\qquad
 \Sigma_{\rm inc}(i\omega_n)=-i\gamma_\infty\Big[1-\frac{\Omega_c^2}{a(\omega_n+a)}\Big],\qquad a=\sqrt{\Omega_c^2+\pi^2T^2}.
 \label{eq:siginc}
\end{equation}
On the Matsubara axis $\Sigma_{\rm inc}$ is purely imaginary. On the real axis causality gives it the real part $-\gamma_\infty\Omega_c^2\omega/[a(\omega^2+a^2)]$, odd in $\omega$, which vanishes at the Fermi level, so $\Sigma_{\rm inc}$ does not shift the level of $F$ there. The fitted $\gamma_\infty$ is comparable to the Hubbard-III rate $(\Nf+1)|\Impart\Delta_f(i\omega_n)|$ at $\omega_n=U/2$. The scattering rate of $\Sigma_{\rm inc}$ at the Fermi level, $\gamma_\infty\pi^2T^2/(\Omega_c^2+\pi^2T^2)$, vanishes as $T^2$. This incoherent part is therefore essentially a constant scattering rate of $F$, cut off below $\Omega_c$.

\subsection{Temperature evolution of the spectra}
\label{app:akwT}\label{app:bands}

\begin{figure*}[t]
 \includegraphics[width=\textwidth]{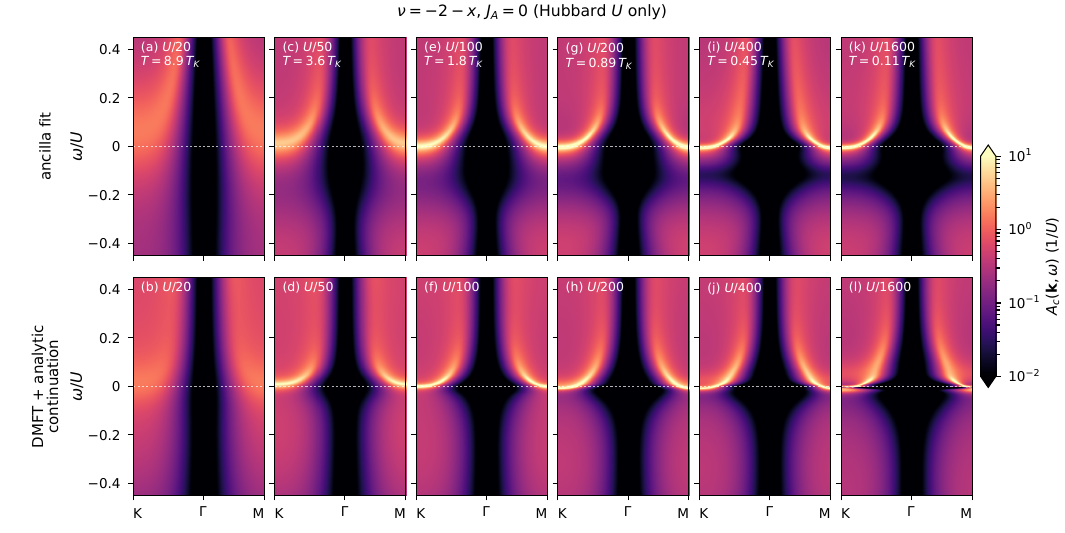}
 \caption{Evolution of the spectra from the bad metal to the heavy Fermi liquid with the Hubbard interaction alone ($J_A=0$, $s^2=0.05$, $x\simeq0.2$ beyond $\nu=-2$). Panels (a, c, e, g, i, k) show $A_c(\bk,\omega)$ along $K$--$\Gamma$--$M$ from the ancilla fit at each temperature, on the real axis without continuation. Panels (b, d, f, h, j, l) show it from DMFT with analytic continuation, the maximum-entropy continuation of the same self-energy. At $U/1600$ the continuation has an artifact at the Fermi level. The color scale is logarithmic, and the dashed line is the Fermi level. }
 \label{fig:chainakw}
\end{figure*}

\begin{figure*}[t]
 \includegraphics[width=\textwidth]{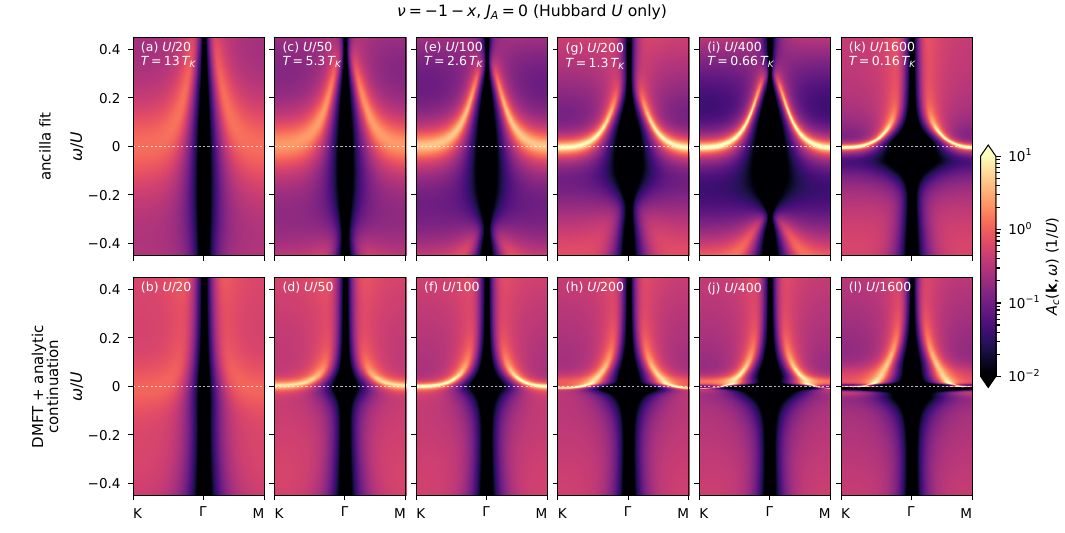}
 \caption{As Fig.~\ref{fig:chainakw}, beyond $\nu=-1$ ($J_A=0$, $s^2=0.05$, $x\simeq0.2$, $T_K=U/263$), with the ancilla fit in (a, c, e, g, i, k) and DMFT with analytic continuation in (b, d, f, h, j, l). At $U/400$ and $U/1600$ the continuation has an artifact at the Fermi level. }
 \label{fig:chainakwodd}
\end{figure*}

We show the temperature evolution of the electron spectrum at $\nu=-2-x$ for the Hubbard interaction alone, $J_A=0$, at $s^2=0.05$, from $U/20$, well above $T_K$, down to $U/1600$ (Fig.~\ref{fig:chainakw}). The analytical fit with the ancilla theory gives $\Sig(\omega)$ on the real axis without analytic continuation, through Eqs.~\eqref{eq:sigcf} and \eqref{eq:siginc}. Its spectra agree with the maximum-entropy continuation of the same DMFT self-energy. Above $T_K$ the heavy band lies above the Fermi level and is broad, a few $k_BT$ wide. On cooling it narrows and comes down, crosses the Fermi level near $T_K$, and below $T_K$ it is the heavy band of the large Fermi surface. At $\nu=-1-x$ the spectra evolve in the same way (Fig.~\ref{fig:chainakwodd}).

The heavy band lies above the Fermi level at finite temperature because of level repulsion. Near the zone boundary the electron sits at $S_{\bk}\Repart\Sig(0)-\mu$, with $\Repart\Sig(0)=\Sigma_\infty-\Phi^2\epsilon^*_F/(\epsilon^{*2}_F+\gamma_F^2)$, where $\epsilon^*_F$ and $\gamma_F$ are the level and the decay rate of $F$ at the Fermi level. The first term, the Hartree energy, puts the electron above the Fermi level. The second term is the repulsion from $F$, which lies far above the Fermi level and pushes the electron down. At $T\ll T_K$, $F$ does not decay at the Fermi level. The spinon gives $\Sigma_F$ a pole, but this pole is sharp and lies below the Fermi level, at $\epsilon_{F'}$. $F$ therefore scatters strongly only near $\epsilon_{F'}$, while at the Fermi level the pole only shifts $F$ up, by $B^2/|\epsilon_{F'}|$. $\Sigma_{\rm inc}$ also vanishes at the Fermi level. The repulsion from $F$ then brings the band just below the Fermi level. On warming, $F$ decays at the Fermi level at a rate comparable to its level, mostly because the dip of $\Sigma_{\rm inc}$ fills in. This weakens the repulsion by about half and lifts the band above the Fermi level. The fall of $\mu$ with temperature at fixed filling lifts it further. The ancilla theory shows this directly [Fig.~\ref{fig:chainbands}(d)]. With the parameters frozen at their values at $U/400$, except those that follow the temperature, the band at $K$ rises above the Fermi level on warming. If $\Sigma_{\rm inc}$ is kept at its $T=0$ form, the band stays at the Fermi level. The heavy band thus lies above the Fermi level because of the scattering of $F$.

\section{Anti-Hund's coupling and the pseudogap metal}
\label{app:pg}

\subsection{Two forms of the anti-Hund's coupling}

The eight flavors are $a=(o,v,s)$, with the orbital $o=\pm$ of angular momentum $L_z=\pm1$ about the AA site, the valley $v=K,K'$ and the spin $s$. We use two forms of the anti-Hund's coupling. The first is $H_A^d$ of Eq.~\eqref{eq:had},
\begin{equation}
 \begin{gathered}
 H_A^d=-J_A^d\sum_{i,o=\pm}\Pi_{io}^\dagger\Pi_{io},\\
 \Pi_o^\dagger=\tfrac{1}{\sqrt2}\big(\hf^\dagger_{oK\uparrow}\hf^\dagger_{oK'\downarrow}-\hf^\dagger_{oK\downarrow}\hf^\dagger_{oK'\uparrow}\big).
 \end{gathered}
 \nonumber
\end{equation}
It lowers the two intra-orbital inter-valley spin singlets with $L_z=\pm2$. The second is the coupling of Ref.~\cite{ZhaoZhang2025},
\begin{equation}
 \begin{gathered}
 H_A=-J_A\sum_i\Pi_i^\dagger\Pi_i,\\
 \Pi^\dagger=\tfrac12\big[f^\dagger_{+K\uparrow}f^\dagger_{-K'\downarrow}-f^\dagger_{+K\downarrow}f^\dagger_{-K'\uparrow}
 +f^\dagger_{-K\uparrow}f^\dagger_{+K'\downarrow}-f^\dagger_{-K\downarrow}f^\dagger_{+K'\uparrow}\big],
 \end{gathered}
 \label{eq:ha}
\end{equation}
which lowers the single inter-orbital inter-valley spin singlet with $L_z=0$. In both forms each flavor $a$ has one partner $\bar a$ in the other valley, and the eight flavors remain equivalent. The filling scans of Figs.~\ref{fig:intro}(c) and \ref{fig:metal}(d,e) use $H_A^d$. Appendices~\ref{app:pgfit} and \ref{app:polaron} and Fig.~\ref{fig:metal}(f) use $H_A$. At equal strength the two forms give a pseudogap of the same size. We solve the impurity with TRIQS/cthyb~\cite{Seth2016,Parcollet2015}.

\subsection{Pseudogap temperature}
\label{app:jad}

\begin{figure}[h]
 \includegraphics[width=\linewidth]{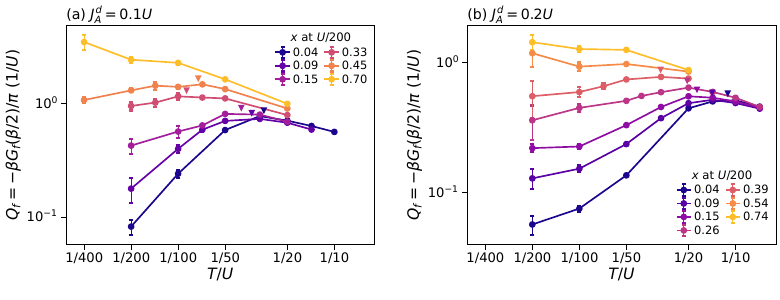}
 \caption{Pseudogap temperature. (a,b) Weight of the orbital within about $2T$ of the Fermi level, $Q_f$ of Eq.~\eqref{eq:qf}, against temperature at fixed $\mu$, for $J_A^d=0.1U$ and $0.2U$. The legend gives the doping at $U/200$. Triangles mark $T^*$, the maximum of $Q_f$. }
 \label{fig:jadtstar}
\end{figure}

We define $T^*$ through the weight of the orbital within about $2T$ of the Fermi level. This weight is local and not resolved in momentum. It comes from the local Green function of the orbital, $G_f(z)=\avg{S_{\bk}G_c(\bk,z)}$ of Eq.~\eqref{eq:dmft}, which sums all Bloch states, each with its orbital weight $S_{\bk}$. It needs no continuation,
\begin{equation}
 Q_f(T)=-\frac{\beta}{\pi}\,G_f(\tau=\beta/2)=\int\dd\omega\,\frac{A_f(\omega)}{2\pi T\cosh(\omega/2T)} .
 \label{eq:qf}
\end{equation}
In a metal $Q_f$ grows on cooling. Once a gap opens at the Fermi level, it falls. $T^*$ is the temperature of its maximum [Fig.~\ref{fig:jadtstar}(a,b)]. 

Near $\nu=-2$, $T^*\simeq0.4J_A^d$ is the temperature at which the singlets form. With doping it falls roughly exponentially, $T^*\propto e^{-x/0.36}$ for $J_A^d=0.1U$, the solid line of Fig.~\ref{fig:intro}(c). At the largest dopings studied, near $x\simeq0.7$, $Q_f$ has no maximum and grows on cooling as in a Fermi liquid. The coupling $H_A$ with $J_A=0.1U$ gives $T^*=U/42$ at $x=0.2$, close to $H_A^d$ of the same strength. Below $T^*$ the electron band splits at the Fermi level [Fig.~\ref{fig:metal}(e)]. At $x\simeq0.15$ and $J_A^d=0.1U$ the edges at $K$ lie at $-0.03U$ and $+0.06U$, a gap of order $J_A^d$. This gap narrows and fills with doping. Near $\nu=-2$ we find the pseudogap only on the hole-doped side.

\begin{figure}[h]
 \includegraphics[width=\linewidth]{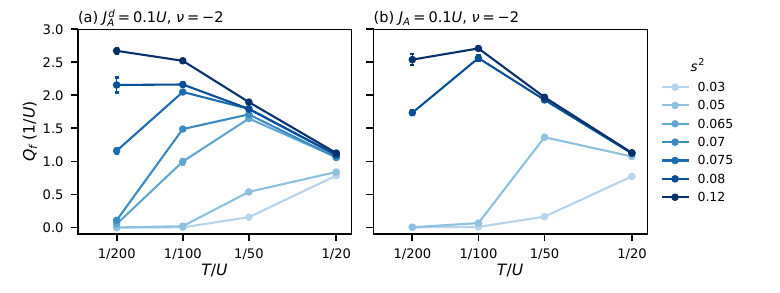}
 \caption{Spin-singlet insulator at $\nu=-2$. Weight of the orbital at the Fermi level, $Q_f$ of Eq.~\eqref{eq:qf}, against temperature at $n_c=2$ for the $s^2$ in the legend, (a) with $J_A^d=0.1U$ of Eq.~\eqref{eq:had} and (b) with $J_A=0.1U$ of Eq.~\eqref{eq:ha}. In the insulator $Q_f$ vanishes on cooling, and in the metal it stays large. }
 \label{fig:nu2ja}
\end{figure}

\emph{Insulator at $\nu=-2$.}---The anti-Hund's coupling also enhances the insulating gap at $\nu=-2$, while it leaves the Mott transitions at $\nu=-1$ and $-3$ unchanged.  At $T=U/200$ the insulating gap survives up to $s^2\simeq0.075$ for $J_A^d=0.1U$ and beyond $s^2=0.05$ for $H_A$, two to three times $s^2_c$ of the Mott insulator without $J_A$ [Fig.~\ref{fig:nu2ja}]. 

\subsection{Pseudogap metal at $x=0.2$ and its analytical fit}
\label{app:pgfit}

\begin{figure*}[t]
 \includegraphics[width=\textwidth]{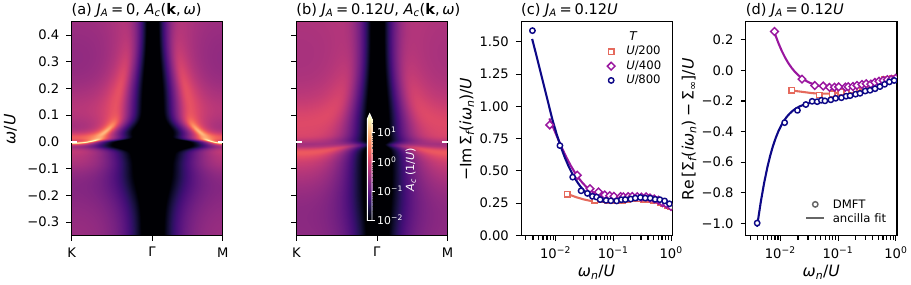}
 \caption{Pseudogap metal with the anti-Hund's coupling of Eq.~\eqref{eq:ha} ($s^2=0.05$, $x=0.2$). (a,b) $A_c(\bk,\omega)$ at $T=U/200$ without it and for $J_A=0.12U$ (maximum entropy, logarithmic color scale). (c) $-\Impart\Sig(i\omega_n)$ and (d) $\Repart[\Sig(i\omega_n)-\Sigma_\infty]$ at $J_A=0.12U$ for the temperatures in the legend. Symbols are DMFT, and lines are the ancilla fit of Eq.~\eqref{eq:pgchain}. }
 \label{fig:pgscan}
\end{figure*}

\emph{Crossover.}---At $T=U/200$ a coupling $J_A\le0.02U$ leaves the heavy Fermi liquid of $J_A=0$ almost unchanged. Between $J_A=0.02U$ and $0.04U$, four to seven times $T_K$, the moments form local singlets that disrupt the Kondo screening. For $J_A\ge0.08U$ the electron scattering rate grows on cooling, and neither the electron nor $F$ has a Fermi surface [Fig.~\ref{fig:pgscan}(a--c)]. On cooling this threshold falls. At $U/400$ the heavy Fermi liquid requires $J_A<0.01U$, and a coupling $J_A>0.01U$ already opens the pseudogap.

\begin{figure}[h]
 \includegraphics[width=\linewidth]{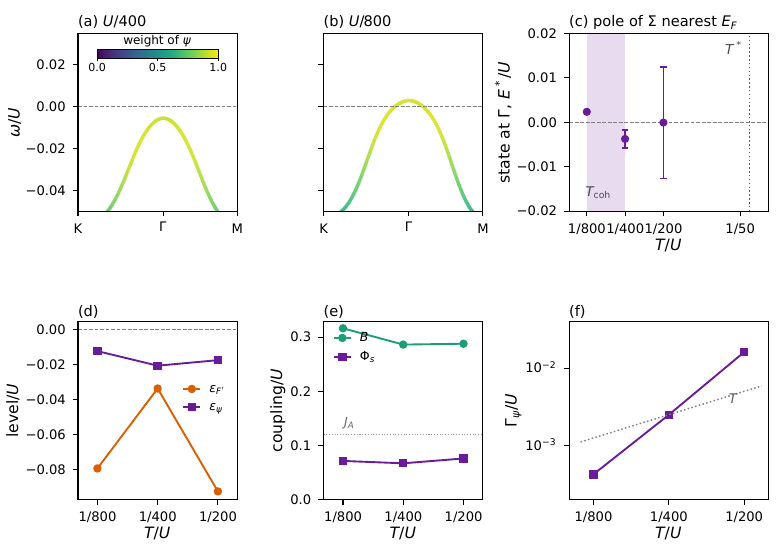}
 \caption{Pseudogap metal in the ancilla theory ($J_A=0.12U$, $s^2=0.05$, $x\simeq0.2$). (a,b) Bands of Eq.~\eqref{eq:hpgchain} at $U/400$ and $U/800$, colored by the weight of $\psi$. (c) Energy of the state at $\Gamma$ against temperature, with its width as bars. (d--f) Fitted levels, couplings and width of $\psi$ against temperature. Dashed lines mark the Fermi level.}
 \label{fig:pgchainT}
\end{figure}

\emph{Ancilla chain with $\psi$.}---The one-pole form of Eq.~\eqref{eq:chain} fails in the pseudogap metal. The singlets gap the spinon $F'$, so the pseudogap is a spinon gap of the heavy Fermi liquid. To capture it we add a second pair of ancilla fermions $(\psi,\psi')$ with the coupling $\Phi_s(F'^\dagger\psi+{\rm H.c.})$. Without damping, the Hamiltonian is
\begin{equation}
 H_{\bk}=\begin{pmatrix} S_{\bk}\Sigma_\infty-\mu & \sqrt{S_{\bk}}\,\Phi & 0 & 0\\ \sqrt{S_{\bk}}\,\Phi & \epsilon_F & B & 0\\ 0 & B & \epsilon_{F'} & \Phi_s\\ 0 & 0 & \Phi_s & \epsilon_\psi\end{pmatrix}
 \label{eq:hpgchain}
\end{equation}
in the basis $(c_{\bk},F_{\bk},F'_{\bk},\psi_{\bk})$. Eliminating $\psi$, $F'$ and $F$ in turn gives the self-energy of the electron,
\begin{equation}
 \Sig(z)=\Sigma_\infty+\cfrac{\Phi^2}{z-\epsilon_F-\Sigma_{\rm inc}(z)-\cfrac{B^2}{z-\epsilon_{F'}-\cfrac{\Phi_s^2}{z-\epsilon_\psi+i\Gamma_\psi}}},
 \label{eq:sigpg}
\end{equation}
which is Eq.~\eqref{eq:pgchain} inserted in Eq.~\eqref{eq:dysonF}. The scattering rate $\Gamma_\psi$ of $\psi$ stands for its coupling to $\psi'$. As in Appendix~\ref{app:spinon}, $\Sigma_\infty$ and $\Phi^2$ are measured. Eight parameters are fitted: the level $\epsilon_F$, the rate $\gamma_\infty$ and cutoff $\Omega_c$ of $\Sigma_{\rm inc}$ [Eq.~\eqref{eq:siginc}], the coupling $B^2$, the spinon level $\epsilon_{F'}$, the coupling $\Phi_s$, the level $\epsilon_\psi$ and the rate $\Gamma_\psi$. This chain reproduces the self-energy to a few percent from $U/200$ to $U/800$ [Fig.~\ref{fig:pgscan}(c,d)]. The anti-Hund's coupling sets the coupling to $\psi$, $\Phi_s\simeq0.6J_A$, and halves the residue $B^2$. Figure~\ref{fig:pgchainT}(d--f) shows how the parameters depend on temperature.

\emph{The state at $\Gamma$.}---At $\Gamma$, where $S_{\bk}=0$, the electron decouples, and the top of the band is the state from $F$--$F'$--$\psi$ nearest the Fermi level. $F$ lies far above the Fermi level and $F'$ far below it, so this state is mostly $\psi$ [Fig.~\ref{fig:pgchainT}(a,b)]. Its energy is the level of $\psi$ pushed up by the repulsion from $F'$ below,
\begin{equation}
 E^*\simeq\epsilon_\psi+\frac{\Phi_s^2}{|\epsilon_{F'}|+B^2/\epsilon_F},
\end{equation}
and its width is set by $\Gamma_\psi$. Near the Fermi level $\Sig\simeq\Sigma_\infty+w/(z-E^*+iW)$, with $w/\Phi^2$ the weight of $F$ in this state and $W$ its width. On the Matsubara axis $\Repart[\Sig-\Sigma_\infty](i\pi T)\simeq-wE^*/[E^{*2}+(\pi T+W)^2]$. It is positive while the state lies below the Fermi level and becomes large and negative once the state lies above it. This sign change leads to the qualitative change of $\Repart\Sig$ between $U/400$ and $U/800$ in Fig.~\ref{fig:pgscan}(d).

\emph{Temperature evolution.}---Above $T^*$ the singlets have not formed, and the state is the bad metal of $J_A=0$. Below $T^*$ the singlets gap the heavy band and disrupt the Kondo screening, and a level $\psi$ near the Fermi level becomes necessary in the fit only near $U/200$, well below $T^*$, which is $U/42$ for $J_A=0.1U$. At higher temperature $\psi$ is not necessary for the fit, so Fig.~\ref{fig:pgchainT} shows its parameters from $U/200$ down. At first $\psi$ is broad. Its width, several times $T$, exceeds its distance from the Fermi level [Fig.~\ref{fig:pgchainT}(c)]. The doped holes then sit in an incoherent level, there is no Fermi surface, and the electron scattering rate grows on cooling. This is the pseudogap metal between $T_{\rm coh}$ and $T^*$. On cooling $\Gamma_\psi$ falls steeply, roughly as $T^{2.6}$, from about $3T$ at $U/200$ to $T$ at $U/400$ and $0.3T$ at $U/800$ [Fig.~\ref{fig:pgchainT}(f)]. Meanwhile $\Phi_s$ stays close to $0.6J_A$ [Fig.~\ref{fig:pgchainT}(e)]. At $U/400$ the fits place the sharp state just below the Fermi level, so the band top at $\Gamma$ is still occupied [Fig.~\ref{fig:pgchainT}(a)]. The band top rises through the Fermi level between $U/400$ and $U/800$ [Fig.~\ref{fig:pgchainT}(b,c)] and a small Fermi surface forms below $T_{\rm coh}$, which gives the sFL phase. 

\subsection{The spin polaron and the small Fermi surface}
\label{app:polaron}

Below $T_{\rm coh}$ the quasiparticles of the small Fermi surface are carried mainly by $\psi$. At $\bk_0$, where $S_{\bk}=0$, the electron has no weight in them, and $F$ has only a small weight. Here we ask which physical operator $\psi$ stands for. In general it is a composite operator that also has non-local parts. We restrict ourselves to on-site operators and look for the one with the largest share in this state.

\emph{Exact diagonalization.}---The Monte Carlo solver cannot measure correlation functions of composite operators. We therefore diagonalize the same impurity at $T=0$, with the converged bath represented by a few levels per flavor. At $\bk_0$, where $\hf_{\bk_0}=0$, the lattice states are the poles of $\Sig$, the zeros of $G_f$. The share of a local operator $A$ in such a pole is the residue of its $f$-irreducible propagator, $\tilde G_{AA}=G_{AA}-G_{Af}G_f^{-1}G_{fA}$, divided by $(A_\perp|A_\perp)$, with $A_\perp=A-(A|f)f$ and the inner product $(A|B)=\langle\{A,B^\dagger\}\rangle$. The pole nearest the Fermi level lies at $-0.019U$, and $F$ has only the share $0.04$ in it. The few bath levels cannot follow the band top across the Fermi level, so this pole stays below it.

\emph{Spin polaron.}---For $a=(+K\uparrow)$ and $\bar a=(-K'\downarrow)$ the commutator of the electron with the anti-Hund's term is
\begin{equation}
 [f_a,H_A]=-\frac{J_A}4\big(\hat O_a+P_a\big),\qquad
 \hat O_a=n_{-K'\downarrow}f_{+K\uparrow}-S^-_{-K'}f_{+K\downarrow}.
\end{equation}
Here $\hat O_a=[(\tfrac12n_{-K'}-\bm S_{-K'}\cdot\bm\sigma)f_{+K}]_\uparrow$ removes the electron from a singlet with the spin of the partner orbital in the other valley. The term $P_a$ transfers a pair from the other singlet channel. The best combination of all local Hubbard operators with the quantum numbers of $f_a$ has the share $0.36$ in the pole nearest the Fermi level. Almost all of this share comes from a single operator. With the singlet doublon $S=\Pi^\dagger|0\rangle$, the triplons $t_b=(2/\sqrt3)f^\dagger_bS$ and the other doublons $d_{bc}=f^\dagger_bf^\dagger_c|0\rangle$~\cite{ZhaoMixed2025}, it is
\begin{equation}
 \psi_a\simeq\sum_{b\ne a,\bar a}|d_{b\bar a}\rangle\langle t_b|+\gamma\,|\bar a\rangle\langle S|,
 \label{eq:psi2}
\end{equation}
with $\gamma=-0.29$, against $\gamma=-\sqrt3/2$ for the bare electron $f_a$ projected onto the same transitions. Its share is $0.33$, against $0.04$ for $F$. Both terms remove the electron $a$ from a singlet with its partner $\bar a$ in the opposite valley. Relative to the bare electron, $\psi$ weights the removal from a triplon three times more. In both terms the removed electron breaks a singlet and leaves the spin of its partner $\bar a$ unpaired. The quasiparticle is therefore an electron bound to the unpaired spin of the other valley, as a spin polaron in a doped antiferromagnet is a carrier bound to the spins it disturbs. The operator $\hat O_a$ that the anti-Hund's term generates has the same form, the electron times the spin of the partner orbital.

\emph{Quasiparticle at $U/800$.}---On cooling, the pole of $\Sig$ near the Fermi level moves up and crosses it between $U/400$ and $U/800$, which sets the coherence temperature $T_{\rm coh}\sim U/800$. At $U/400$ the band top at $\bk_0$ still lies just below the Fermi level, and the weight $Q_\beta(S)$ of $F$ peaks there. At $U/800$ and $x=0.23$ it has a ring that encloses $0.27\pm0.05$ holes, a small Fermi surface of about $x$ holes [Fig.~\ref{fig:metal}(f)]. The fitted ancilla theory puts the state at $\bk_0$ at $+0.0024U$, just above the Fermi level, with the share $0.89$ of $\psi$.  The quasiparticle of the small Fermi surface is thus the $\psi$ fermion.

\section{Inter-valley coherence}
\label{app:ivc}

The pseudogap metal above does not break any symmetry, though it violates the perturbative Luttinger theorem. A pseudogap and a small Fermi surface could also come from a broken symmetry, such as IVC order. Here we allow IVC order in DMFT and follow its onset. We show that IVC sets in below $T^*$ and is thus not essential for the pseudogap. Instead it emerges within the symmetric pseudogap phase.

\begin{figure}[h]
 \includegraphics[width=\linewidth]{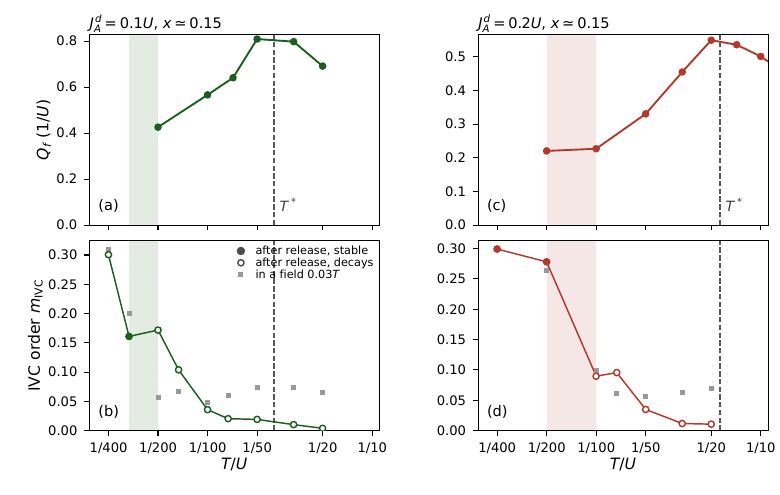}
 \caption{The pseudogap opens before inter-valley coherence sets in ($x\simeq0.15$, $s^2=0.05$). (a,c) Weight of the orbital at the Fermi level, $Q_f$ of Eq.~\eqref{eq:qf}, of the symmetric state against temperature for $J_A^d=0.1U$ and $0.2U$. The dashed line marks $T^*$, the maximum of $Q_f$. (b,d) IVC order $m_{\rm IVC}=\sum_a(\tau_zs_z)_{aa}n_a$, which the Sp(4) symmetry of $H_A^d$ maps onto the spin-singlet IVC. Circles give the order after a kick and a release without field. They are filled where the order is stable and open where it still decreases during the iterations. Gray squares give the response to a field of $0.03T$, applied on cooling from the symmetric state. The shaded band brackets the onset of the order.}
 \label{fig:ivcT}
\end{figure}

The coupling $H_A^d$ of Eq.~\eqref{eq:had} commutes with an Sp(4) rotation of the four flavors of each orbital. This rotation maps the spin-singlet inter-valley coherence $\sigma_0\tau_xs_0$ onto the spin-valley locked order $\sigma_0\tau_zs_z$, with $\sigma$, $\tau$ and $s$ the orbital, valley and spin. We therefore allow IVC through the latter order, which is diagonal in the flavors. Figure~\ref{fig:ivcT} shows how the order appears on cooling at $x\simeq0.15$. $Q_f$ peaks at $T^*$, where the pseudogap opens, and falls below it. The IVC order sets in between $U/300$ and $U/200$ for $J_A^d=0.1U$ and between $U/200$ and $U/100$ for $0.2U$, five to ten times below $T^*$. At $x\simeq0.47$ and $J_A^d=0.1U$ it sets in between $U/100$ and $U/75$, close to $T^*=U/74$.

We decide the onset from two runs at each chemical potential and temperature. One starts from a polarized state and is then released without a field. The other starts from the symmetric state in a weak field of $0.03T$ and is cooled step by step. The order is stable where both runs reach the same nonzero value. It decays where the released order vanishes and the field induces only a small response. Because the iterations converge slowly near the transition, we bracket the onset between the highest temperature at which the order is stable and the next one, at which it decays.

For both strengths of $J_A^d$, IVC sets in below $T^*$ at every doping. For $x\lesssim0.2$ the onset lies at least five times below $T^*$. The two scales approach each other only near $x\simeq0.4$--$0.5$, where the pseudogap is weakest. The pseudogap therefore forms without the order, which develops inside the pseudogap metal. The singlet insulator at $\nu=-2$ has no IVC instability down to $U/400$. 

\end{document}